\documentclass{cernphprep}
                           
\usepackage{graphicx} 
\usepackage{verbatim} 
\usepackage{color}

\newcommand{\npt}{$p_T$}
\newcommand{\npart}{$N_{\rm part}$}
\newcommand{\ncoll}{$N_{\rm coll}$}
\newcommand{\nGc}{GeV/$c$}
\newcommand{\Jpsi} {J/$\psi$}
\newcommand{\npi}{$\pi$}
\newcommand{\nL}{$\Lambda$}
\newcommand{\Kz}{K$^0_{\mathrm S} $}
\newcommand{\nY}{$\Upsilon$}
\newcommand{\dndh}{${\rm d}N_{\rm ch}/{\rm d}\eta$}
\newcommand{\dndy}{${\rm d}N_{\rm ch}/{\rm dy}$}

\newcommand {\PT}          {\ensuremath{p_T}}

\newcommand{\RAA}        {\ensuremath{R_{AA}}}

\begin{document}

\input epsf.tex    
\input epsf.def   

\input psfig.sty


\begin{titlepage}
\PHnumber{2012--005}
\PHdate{15 February 2012}

\title{First Results from Pb+Pb collisions at the LHC}




\begin{abstract}
At the end of 2010, the CERN Large Hadron Collider started operation with heavy ion beams, colliding lead nuclei at a centre-of-mass energy of 2.76 TeV/nucleon and opening a new era in ultra-relativistic heavy ion physics at energies exceeding previous accelerators by more than an order of magnitude. This review summarizes the results from the first year of heavy ion physics at LHC obtained by the three experiments participating in the heavy ion program, ALICE, ATLAS, and CMS.
\end{abstract}

\vfill

\begin{Authlist}
Berndt M\"uller
\Instfoot{a1}{Department of Physics, Duke University, Durham, NC 27708, USA}
J\"urgen Schukraft
\Instfoot{a2}{PH Division, CERN, CH-1211 Geneva 23, Switzerland}
Boles{\l}aw Wys{\l}ouch
\Instfoot{a3}{Department of Physics, Massachusetts Institute of Technology, Cambridge, MA 02139, USA}
\end{Authlist}
\vfill
\Submitted{(To appear in Annual Review of Nuclear and Particle Science)}

\end{titlepage}

\section{Introduction}

After almost two decades of design and construction, the LHC accelerator~\cite{Evans:2008zzb} started operation at the end of 2009 with proton-proton (pp) collisions at 0.9 and 2.36 TeV centre-of-mass energy, reaching its current top energy of $\sqrt{s} = 7$ TeV in 2010. Heavy ion collisions (A+A) are an integral part of the LHC physics program. The first four-week run with Pb beams took place in November 2010 at a centre-of-mass energy of 2.76 TeV per nucleon. About $10~\mu b^{-1}$ of integrated luminosity were delivered to each of the three LHC experiments which take part in the heavy ion program, the two general purpose detectors ATLAS~\cite{Aad:2008zzm} and CMS~\cite{Adolphi:2008zzk} as well as the dedicated heavy ion detector ALICE~\cite{ Aamodt:2008zz,Fabjan:2011jb}. This review summarises the results from the first year of ion physics at the LHC, focusing on global event properties, high \npt~and heavy quark physics, and the production of hidden heavy flavour mesons (\Jpsi, \nY).

\section{The Physics of Relativistic Heavy Ion Collisions}

Relativistic heavy ion collisions make it possible to study the properties of strongly interacting matter at energy densities far above those of normal nuclear matter. General arguments based on the property of asymptotic freedom suggest, and quantitative calculations of lattice quantum chromodynamics confirm \cite{Aoki:2006br,Borsanyi:2010cj,Bazavov:2011nk}, that QCD matter undergoes a transition from a hadronic gas to a quark-gluon plasma at a temperature $T_c \approx 160$ MeV, corresponding to an energy density of $\varepsilon_c \approx 0.5$ GeV/fm$^3$. At small net baryon density, the transition is a smooth cross-over spanning a temperature range of $20-30$ MeV, which means that the precise value of the pseudo-critical temperature $T_c$ depends on the observable used to locate it. The quark-gluon plasma phase is characterized by a much reduced condensate of light quarks, reflecting the approximate restoration of chiral symmetry, and by screening of the chromo-electric force between heavy quarks, implying the absence of quark confinement. At high net baryon density, the confined and deconfined phases are thought to be separated by a first-order phase transition, beginning with a critical point whose position  is not accurately known.

A large body of experimental and theoretical research conducted over the past decades at AGS, SPS, and RHIC has led to the following ``standard model'' of heavy ion collisions at collider energies \cite{Antinori:2003hw,Muller:2006ee}. The energy deposited in the mid-rapidity kinematic range is controlled by the density of gluons contained in the colliding nuclei at moderately small values of the Bjorken scaling variable $x$ \cite{Banerjee:2008nt}. At top RHIC collision energy the relevant range is $\langle x \rangle \approx 10^{-2}$, while at the current LHC energy the value is $\langle x \rangle \approx 10^{-3}$. In this range of $x$ the gluon density obeys a highly nonlinear evolution equation that describes the saturation of the rapidly growing perturbative gluon distribution at low virtuality \cite{Kovchegov:2010pw}.  The virtuality scale at which this saturation occurs is called the saturation scale $Q_s$. At present collider energies and for heavy nuclei $Q_s^2 \approx 1.5-2$ GeV$^2$ at RHIC and $Q_s^2 \approx 3-4$ GeV$^2$ at LHC \cite{Albacete:2011fw}. 

The liberation of the dense gluon sea by interactions during the collision of two nuclei is thought to result in the formation of a dense, non-thermal QCD plasma with highly occupied gauge field modes, often called a ``glasma'' \cite{Lappi:2006fp}. The glasma thermalizes rapidly through nonlinear interactions including plasma instabilities \cite{Romatschke:2006nk,Kurkela:2011ub}. During the longest phase of the heavy ion reaction the QCD matter forms a nearly thermal quark-gluon plasma, whose evolution can be described accurately by relativistic viscous hydrodynamics because of its very small kinematic shear viscosity. The temperature of the plasma at the time of thermalization is approximately $2T_c$; it cools by mainly longitudinal expansion until it converts to a gas of hadron resonances when its temperature falls below $T_c$. At that temperature the chemical composition of the produced hadrons gets frozen, but the spectral distribution of the hadrons is still modified by final-state interactions, which are modelled by a Boltzmann transport equation. 

The very small dimensionless ratio of the shear viscosity $\eta$ to the entropy density $s$ of the QCD matter produced in heavy ion collisions ($4\pi\eta/s \leq 2.5$~\cite{Song:2010mg}) was one of the main discoveries of the experimental program at RHIC \cite{Muller:2006ee}.  Such a small ratio requires that the interactions among the constituent quanta are extremely strong and the cross sections among them approach the unitarity limit \cite{Gyulassy:2004zy}. This picture is supported by QCD lattice simulations, which show that the QCD matter in the temperature range $T_c \leq T \leq 2T_c$ is highly nonperturbative as witnessed, e.~g., by the large value of the trace anomaly of the energy-momentum tensor, $T_\mu^\mu = \varepsilon - 3P$ \cite{Borsanyi:2010cj}.

The strongly coupled nature of the quark-gluon plasma created in the heavy ion reaction is confirmed by its ability to ``quench'' jets. At RHIC, the two main observables revealing this property are the suppression of single hadron yields at high transverse momentum \npt~ relative to $pp$ reactions and the additional suppression of back-to-back emission of high-\npt~hadrons. The ratios quantifying these suppressions are known as $R_{AA}$ and $I_{AA}$, respectively. Compelling arguments related the suppression effects to the ability of the quark-gluon plasma to degrade the kinetic energy of a hard parton (quark or gluon) traversing it, either by elastic collisions or  collisions followed by gluon bremsstrahlung. The large values of the energy loss rate of hard partons deduced from the RHIC data \cite{Bass:2008rv} are consistent with the low shear viscosity deduced from the collective flow pattern of low-momentum hadrons \cite{Majumder:2007zh}.

Before the LHC turned on, it was not clear whether this quantitative model of the dynamics of relativistic heavy ion collisions and the properties of hot QCD matter smoothly extrapolates from RHIC to the one order of magnitude higher LHC energies. Questions of particular interest were whether the increase in the initial temperature of the quark-gluon plasma would result in a much more viscous expansion of the created matter, and whether the gluon saturation effects in the colliding nuclei would be so strong that they significantly reduce, not only the expected density of the glasma, but also the yield of particles at moderately large \npt. Because of the much higher beam energy and the large acceptance of the LHC detectors, heavy ion collisions at the LHC were also expected to provide access to new observables measuring the pattern of jet quenching and extend our knowledge of the interaction of heavy quarks and heavy quarkonia with hot QCD matter. An extensive compilation of theoretical predictions made before the commencement of the LHC heavy ion program can be found in \cite{Abreu:2007kv}.

\section{Global event properties}

Global event properties describe the state and dynamical evolution of the bulk matter created in a heavy ion collision by measuring characteristics of the vast majority of particles which have momenta below a few \nGc, referred to as ``soft'' particles. They include multiplicity distributions, which can be related to the initial energy density reached during the collision, yields and momentum spectra of identified particles, which are determined by the conditions at and shortly after hadronization, and correlations between particles which measure both size and lifetime of the dense matter state as well as some of its transport properties via collective flow phenomena.

\subsection{Multiplicity distributions}

The most basic quantity, and indeed the one measured within days of the first ion collision, is the number of charged particles produced per unit of (pseudo)rapidity, \dndy~(\dndh), in a central, ``head-on'' collision. When the LHC heavy ion program was conceived and detectors had to be designed more than a decade ago, predictions for \dndh~were highly uncertain, ranging from below 1000 to above 4500, because experimental results had to be extrapolated from light ions at low energy (Sulphur beams at $\sqrt{s} = 20$ GeV) by orders of magnitude in both ion mass and beam energy~\cite{JarlskogSatz:1990dv,JarlskogHeinz:1990dv}. With results from RHIC, the uncertainties where substantially reduced, with most predictions concentrating in the range $dN_{\rm ch}/d\eta =1000 - 1700$~\cite{Abreu:2007kv}. The value finally measured at LHC, \dndh~ $\approx 1600$~\cite{Aamodt:2010pb}, was on the high side of this range. From the measured multiplicity one can derive a rough estimate of the energy density with the help of a formula first proposed by Bjorken~\cite{Bjorken:1982qr} relating the energy density to the transverse energy:
\begin{equation}
\varepsilon \geq \frac{dE_T/d\eta}{\tau_0\,\pi R^2} 
= \frac{3}{2} \langle E_T/N\rangle \frac{dN_{\rm ch}/d\eta}{\tau_0\,\pi R^2}
\end{equation}
where $\tau_0$ denotes the thermalization time, $R$ is the nuclear radius, and $E_T/N \approx 1$ GeV is the transverse energy per emitted particle. The value measured at the LHC implies that the initial energy density (at $\tau_0 = 1$ fm/c) is about 15 GeV/fm$^3$~\cite{Krajczar:2011zz}, approximately a factor three higher than in Au+Au collisions at the top energy of RHIC~\cite{Back:2004je,Arsene:2004fa,Adcox:2004mh,Adams:2005dq,Muller:2006ee}. The corresponding initial temperature increases by at least 30\% to $T \approx 300$ MeV, even with the conservative assumption that the formation time $\tau_0$, when thermal equilibrium is first established, remains the same as at RHIC.

The high multiplicity at LHC, together with the large experimental acceptance of the detectors, allow for a very precise determination of the collision geometry (impact parameter and reaction plane orientation) in each event. Events are classified according to ``centrality'', where e.g. the 0-5\% centrality bin contains the 5\% of all hadronic interaction events with the largest final-state multiplicity and therefore the smallest impact parameters. Correspondingly, the  90-100\% centrality bin comprises the 10\% of events with the smallest final-state multiplicity and the largest impact parameters. The experimentally measured centrality can be related analytically or via a probabilistic Glauber Monte-Carlo generator to the number of participating nucleons \npart~which are contained in the nuclear overlap volume for a given centrality class and therefore take part in the collision~\cite{Miller:2007ri}.  Likewise one can extract the number of nucleon-nucleon collisions \ncoll~between the participants. For Pb+Pb at LHC, the centrality or \npart~resolution varies from about 0.5\% for central (\npart~$ \approx 400$) to about 5\% for peripheral (\npart~$< 10$) events~\cite{Aamodt:2010cz,ATLAS:2011ag,Chatrchyan:2011pb}.

\begin{figure}[!t]
\begin{tabular}{cc}
\begin{minipage}{.48\textwidth}
\centerline{\includegraphics[width=1.0\textwidth]{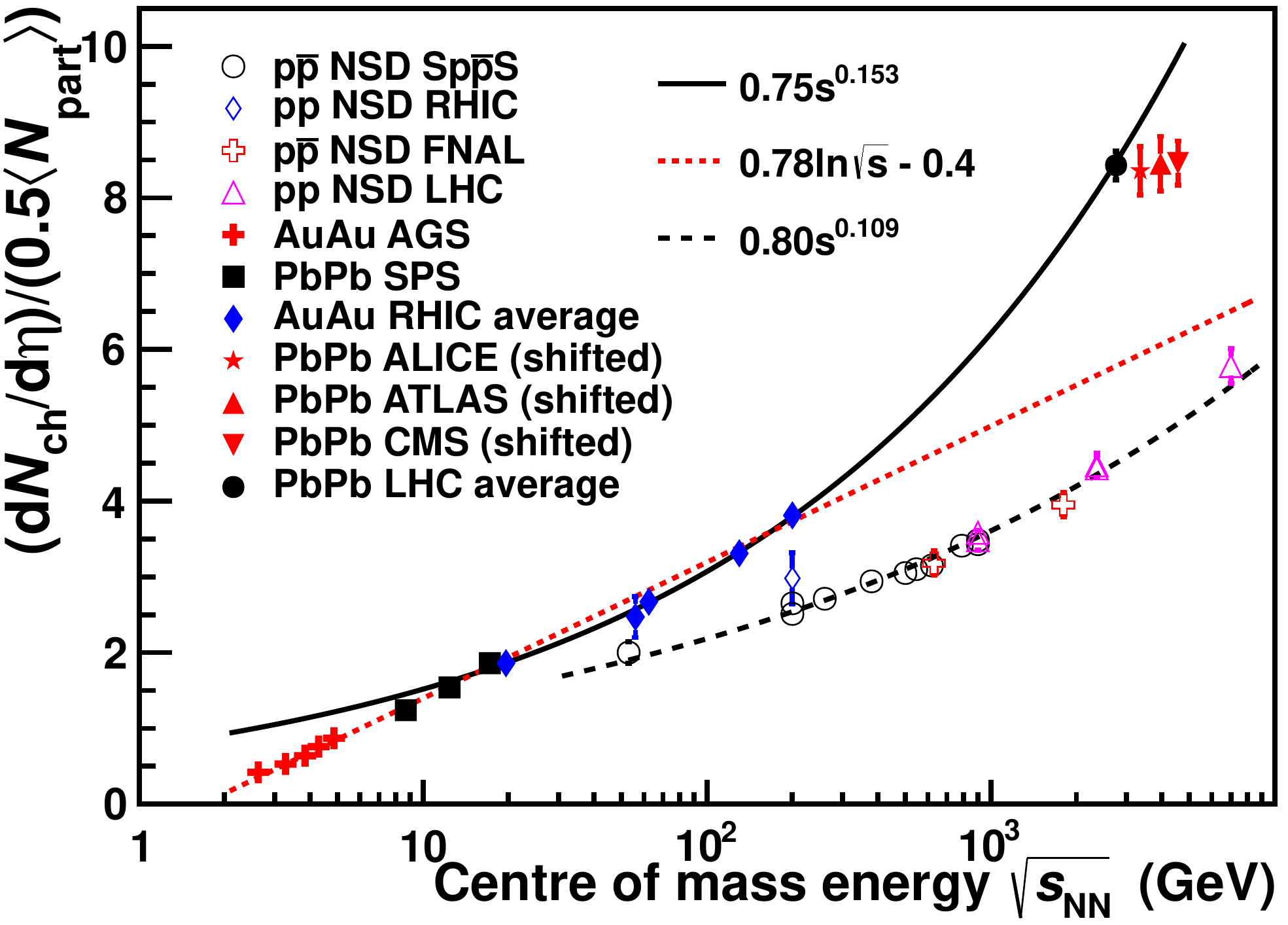}}
\end{minipage} & \begin{minipage}{.48\textwidth}
\centerline{\includegraphics[width=1.0\textwidth]{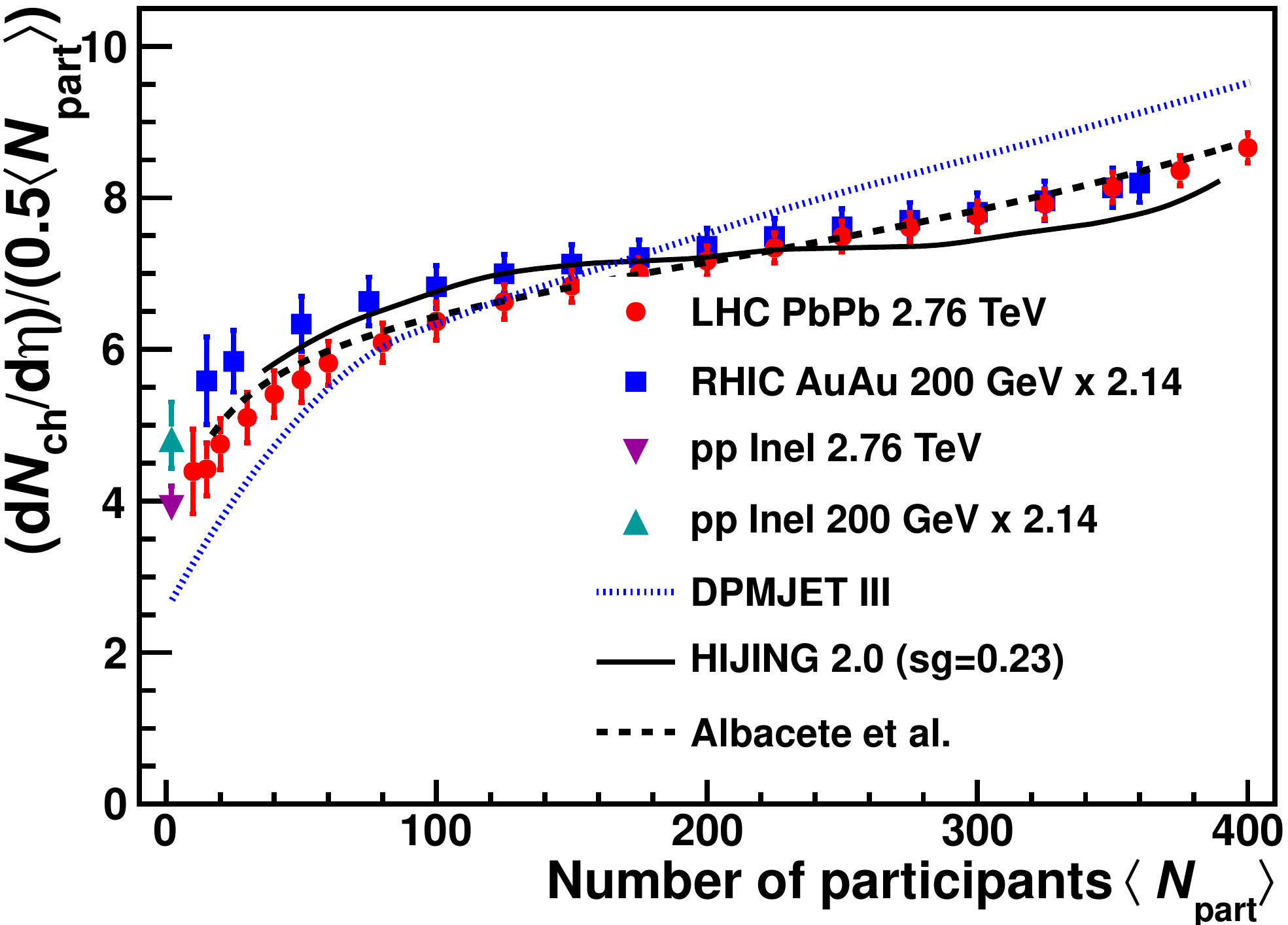}}
\end{minipage}
\end{tabular}
\caption{a): Charged particle pseudorapidity density ${\rm d}N_{\rm ch}/{\rm d}\eta$ per colliding nucleon pair ($0.5 N_{\rm part}$) versus centre of mass energy for $pp$ and A+A collisions. b): ${\rm d}N_{\rm ch}/{\rm d}\eta$ per colliding nucleon pair versus the number of participating nucleons together with model predictions for Pb+Pb at 2.67 TeV
}
\label{JSMult}
\end{figure}

The charged particle multiplicity per participant pair~\cite{Aamodt:2010cz,ATLAS:2011ag,Chatrchyan:2011pb}, \dndh/(0.5 \npart), is shown in Fig.~\ref{JSMult} together with lower energy data~\cite{Adler:2004zn,Abelev:2008ez,Alver:2010ck}  for central A+A collisions (typically 0-5\% or 0-6\% centrality). Note, however, that even for equal centrality selections, the average number of participants or collisions (\npart, \ncoll) are different at the various energies because of different nuclear size (Pb or Au) and the rising $pp$ cross-section. Particle production is no longer compatible with a logarithmic dependence with $\sqrt s$, as it was true for the data up to top RHIC energy~\cite{Alver:2010ck}, but follows a power law $\approx s^{0.15}$. Also the $pp$ data are well described by a power law, however with a less steep dependence on energy ($\approx s^{0.11}$).

The centrality dependence of particle production is compared in Fig.~\ref{JSMult}b with the one measured at RHIC, which is normalised to the LHC result at \npart = 350 by scaling it with a factor 2.14. The results from the three LHC detectors~\cite{Aamodt:2010cz,ATLAS:2011ag,Chatrchyan:2011pb} are in excellent agreement with each other (within 1-2\%) and have been averaged in this figure using the prescription of~\cite{Adler:2004zn}, assuming conservatively that the systematic errors in \npart~are fully correlated between experiments. Comparison to the averaged and scaled 200 GeV Au+Au data (from~\cite{Adler:2004zn}, updated using more recent STAR~\cite{Abelev:2008ez} and PHOBOS~\cite{Alver:2010ck} data) shows a remarkable similarity in the shape of both distributions. For peripheral collisions, however, both distributions extrapolate towards respective values measured in $pp$ inelastic collisions (\npart = 2) at 200 GeV and 2.76 TeV and therefore start to separate because of the different energy dependence seen for $pp$ and A+A in Fig.~\ref{JSMult}a. 

The fact that the shape of the normalised multiplicity distribution varies little with energy had already been noticed at RHIC~\cite{Alver:2010ck}. It was still a surprise that it stays almost constant up to TeV energies, because hard processes, which scale with the number of binary collisions \ncoll, could be expected to contribute significantly to particle production at LHC and lead to a steeper centrality dependence, as predicted by the two component (soft + hard) Dual Parton Model DPMJET~\cite{Bopp:2007sa} (dotted line in Fig.~\ref{JSMult}b). However, a strong impact parameter dependent shadowing of the nuclear parton distribution function can limit this rise with centrality and is responsible for the flatter shape seen in the two component model HIJING~\cite{Deng:2010xg}, which agrees better with the data (full line). Saturation physics based on the ``Colour Glass Condensate'' (CGC) description~\cite{Gelis:2010nm}, an example~\cite{Albacete:2010ad} is shown by the dashed line in Fig.~\ref{JSMult}b,  naturally predicts such a strong nuclear modification, as well as the strong rise of particle production as a function of $\sqrt s$ seen in Fig.~\ref{JSMult}a. In saturation models, the energy dependence is predicted to be a power law with roughly the correct exponent, whereas the centrality dependence changes only weakly (logarithmically) with the saturation scale and therefore the beam energy.

\subsection{Identified particle spectra}

Particle production (\npi, K, p, \nL,..) is a non-perturbative process and cannot be calculated directly from first principles (QCD). In phenomenological QCD inspired event generators, particle spectra and ratios are adjusted to the data in elementary collisions ($pp$, $e^{+}e^{-}$) using a large number of parameters. In heavy ion reactions, however, inclusive particle ratios and spectra at low transverse momentum, which include the large majority of all produced hadrons even at LHC energies (about 95\% of all particles are below 1.5 \nGc), are consistent with simple descriptions by statistical/thermal~\cite{BraunMunzinger:2003zd,Becattini:2009sc} and hydrodynamical~\cite{Huovinen:2006jp} models, where particle ratios are determined during hadronisation at or close to the QGP phase boundary (``chemical freeze-out'', see below), whereas particle spectra reflect the conditions somewhat later in the collision, during ``kinetic freeze-out''.

In $pp$ collisions at high energy, the transverse momentum distribution  can be separated into an exponentially decreasing low \npt~part ($<$ few \nGc), arising from soft, nonpertubative processes, and a powerlaw high \npt~tail from hard QCD scattering and fragmentation.
A defining characteristic of heavy ion collisions is the appearance of ordered motion amongst the emitted hadrons in the soft part of the momentum spectrum~\cite{Voloshin:2008dg,Snellings:2011sz}. It is called collective flow and implies, in contrast to random thermal motion, a strong correlation between position and momentum variables (nearby particles have similar velocities in both magnitude and direction). Flow arises in a strongly interacting medium in the presence of local pressure  gradients. Different flow patterns are observed in heavy ion collisions and classified in terms of their azimuthal angle $\varphi$ dependence with respect to the reaction plane, i.~e.\ the plane spanned by the beam direction and the impact parameter. The isotropic (or angle averaged) component is called {\em radial flow}. In the framework of hydrodynamic models, the fluid properties (viscosity, equation-of-state, speed of sound, ..) together with boundary conditions both in the initial state (collision geometry, pressure gradients, ..) and in the final state (freeze-out conditions) determine the pattern of collective motions and the resulting momentum spectra ${\rm d}^2{\rm N}/{\rm d}p_T{\rm d}\varphi$.

\begin{figure}[!t]
\begin{tabular}{cc}
\begin{minipage}{.48\textwidth}
\centerline{\includegraphics[width=1.0\textwidth]{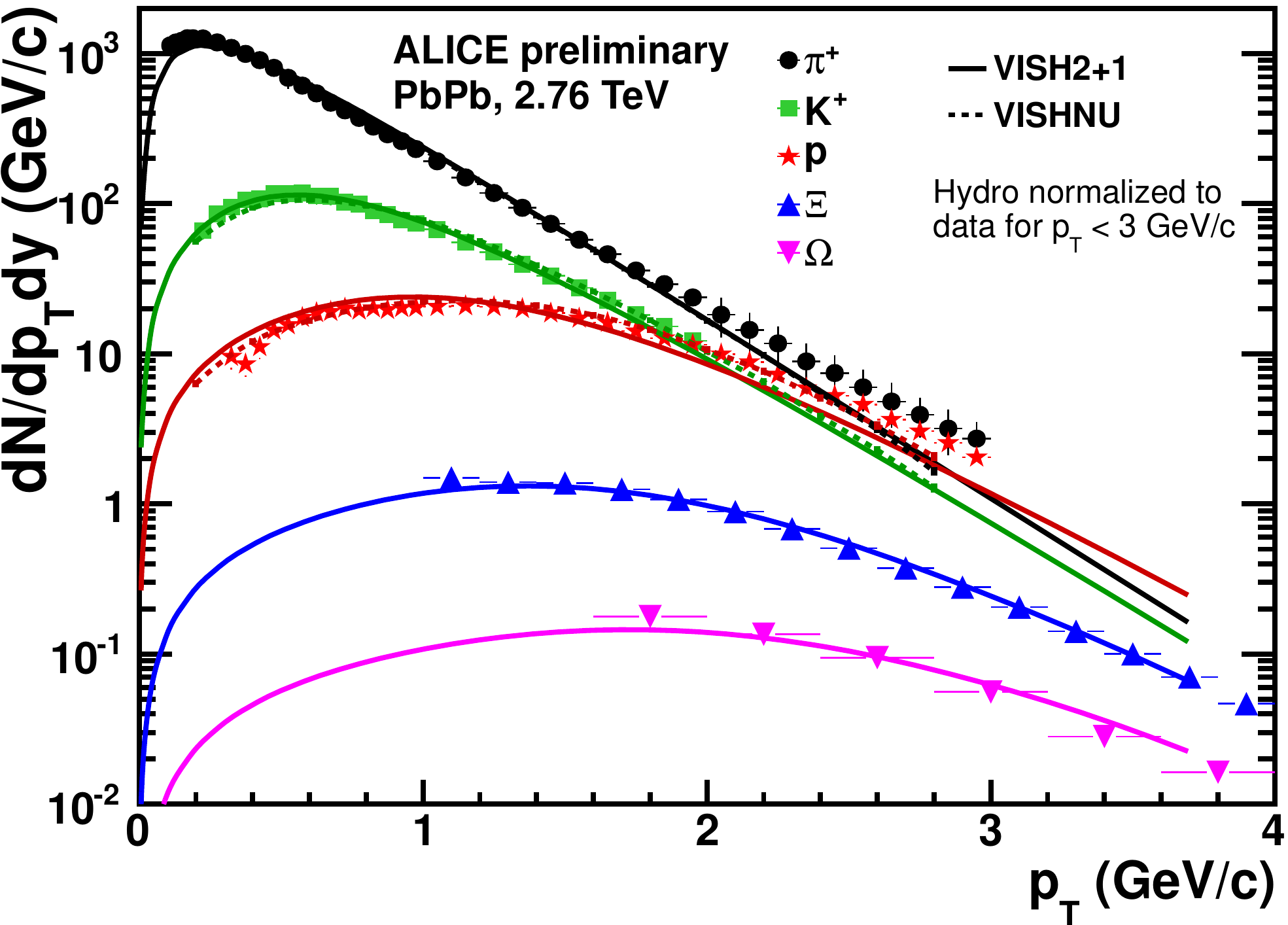}}
\end{minipage} & \begin{minipage}{.48\textwidth}
\centerline{\includegraphics[width=1.0\textwidth]{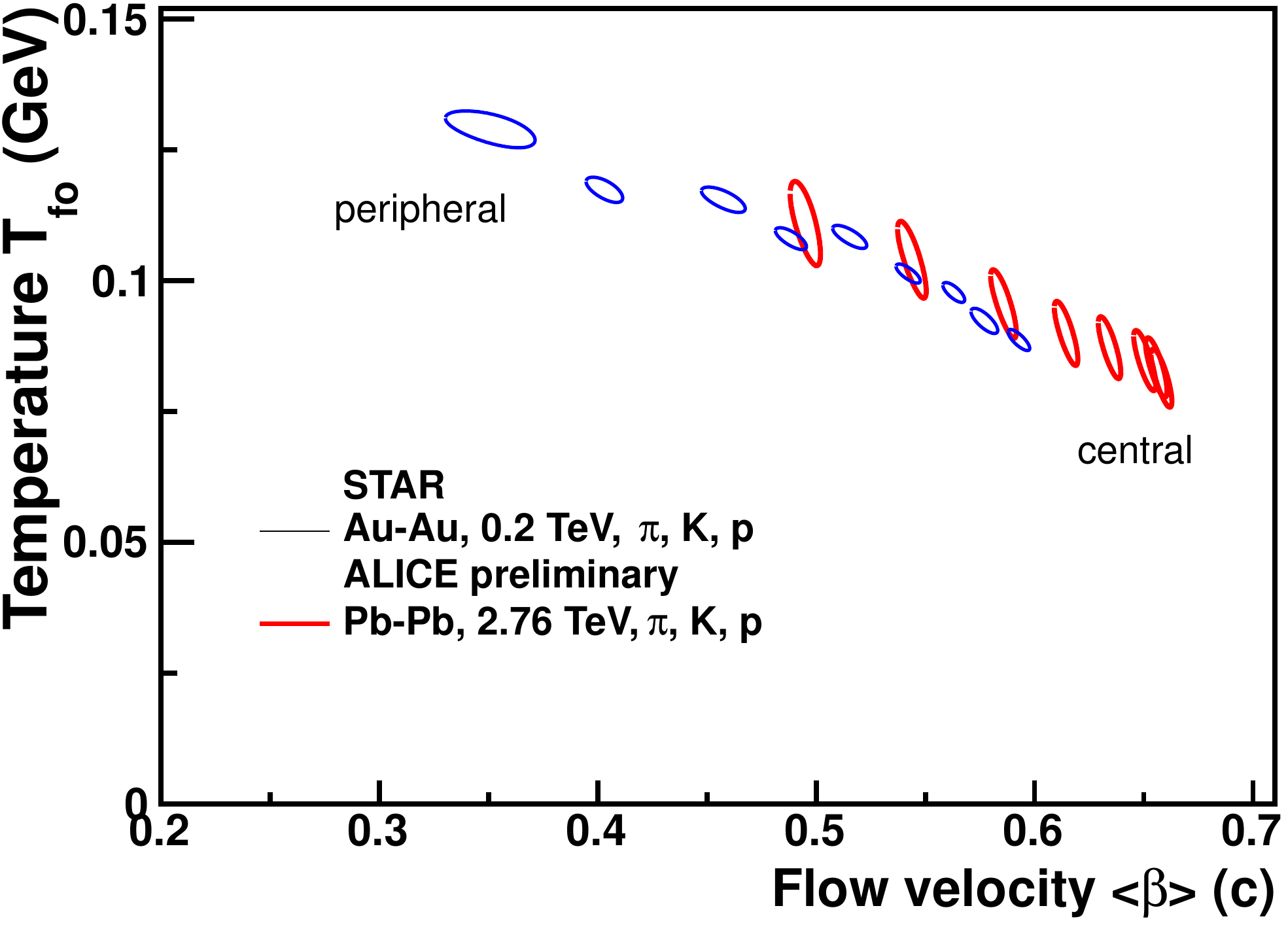}}
\end{minipage}
\end{tabular}

\caption{ a): Transverse momentum spectra of identified particles for central Pb+Pb collisions. The lines are scaled results from a boost invariant hydrodynamic model with (VISHNU, dashed lines) and without (VISH2+1, full lines) rescattering in the hadronic phase. b): Kinetic freeze-out temperatures $T_{fo}$ and average radial flow velocities $\langle\beta\rangle$ extracted from identified particle spectra (\npi, K, p) at LHC (Pb+Pb) and RHIC (Au+Au) for different centralities.
}
\label{JSSpect}
\end{figure}

Fig.~\ref{JSSpect}a shows the transverse momentum distributions of identified particles in central Pb+Pb collisions at the LHC~\cite{Floris:2011ru,MNSQM2011}. The spectral shapes differ significantly from both $pp$ at LHC and Au+Au at RHIC, most dramatically for protons at low \npt. The characteristic mass-dependent blue shift generated by the radial flow leads to a strong depletion at low \npt~-- the proton spectrum is almost flat between 0.5 and 1.5 \nGc~-- and to a harder spectrum at high \npt, where the p/\npi~ratio reaches a value of close to unity around 3 \nGc.

The data are compared to hydrodynamic calculations, which are normalised to the data individually for each particle species in order to better compare the shapes (absolute particle yields and ratios, which are an external input to the hydro models, are discussed below). The dashed line (VISH2+1) is a pure hydro calculation~\cite{Song:2007fn,Song:2007ux,Song:2008si} for $\Xi$ and $\Omega$ (0-20\% centrality); the full line (VISHNU) for \npi, K, p (0-5\% centrality) includes additional final-state rescattering, calculated with the URQMD transport code, which increases the radial flow velocity and improves the agreement with the data~\cite{Song:2010aq}. The hydro model describes the data fairly well up to intermediate momenta. Around 2 \nGc, first pions and later protons start to deviate from the exponential hydro slope, indicating a progressive decoupling of high momentum particles from the thermalized and flowing bulk matter.

\begin{figure}[!t]
\begin{tabular}{cc}
\begin{minipage}{.48\textwidth}
\centerline{\includegraphics[width=1.0\textwidth]{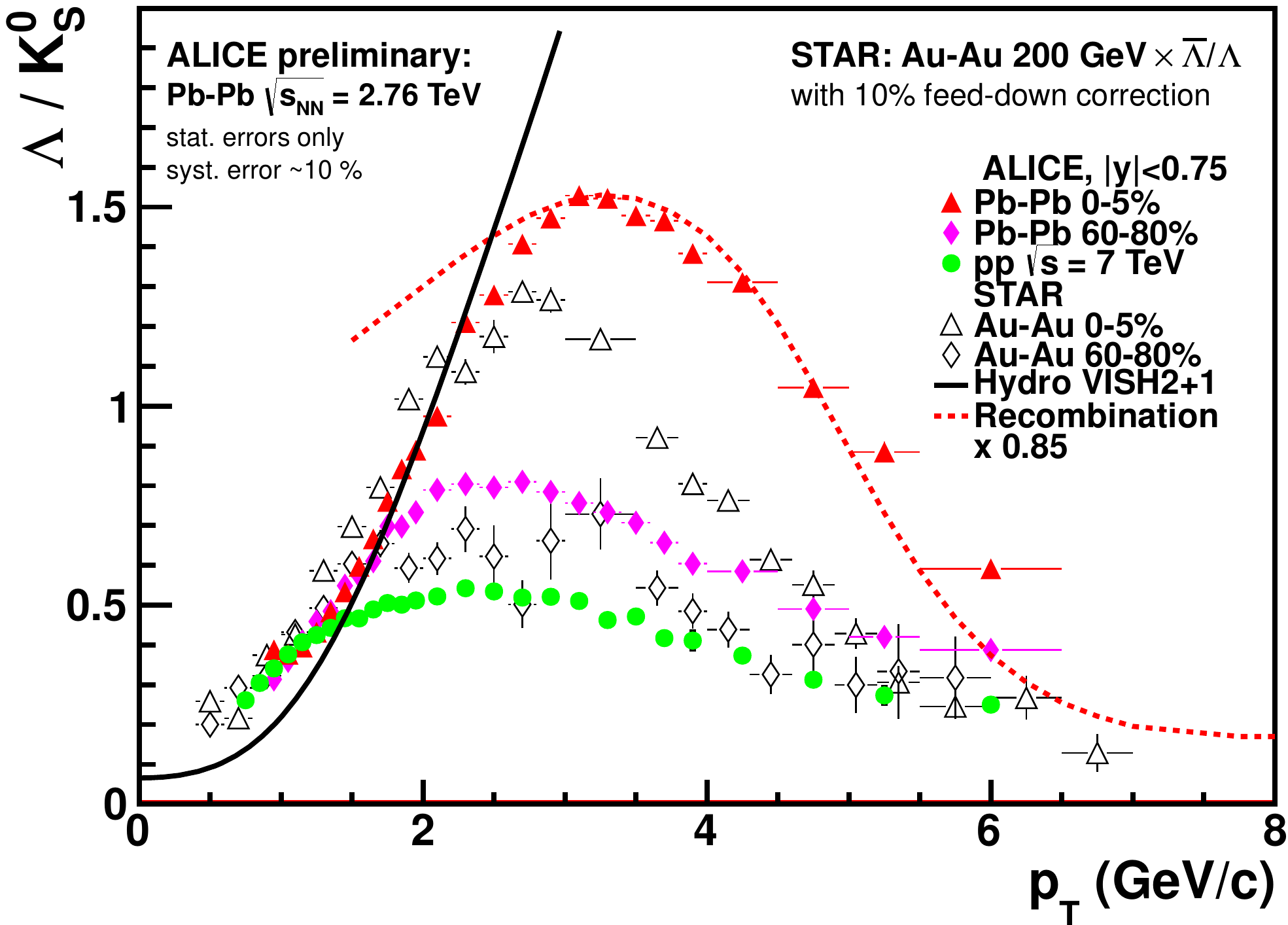}}
\end{minipage} & \begin{minipage}{.48\textwidth}
\centerline{\includegraphics[width=1.0\textwidth]{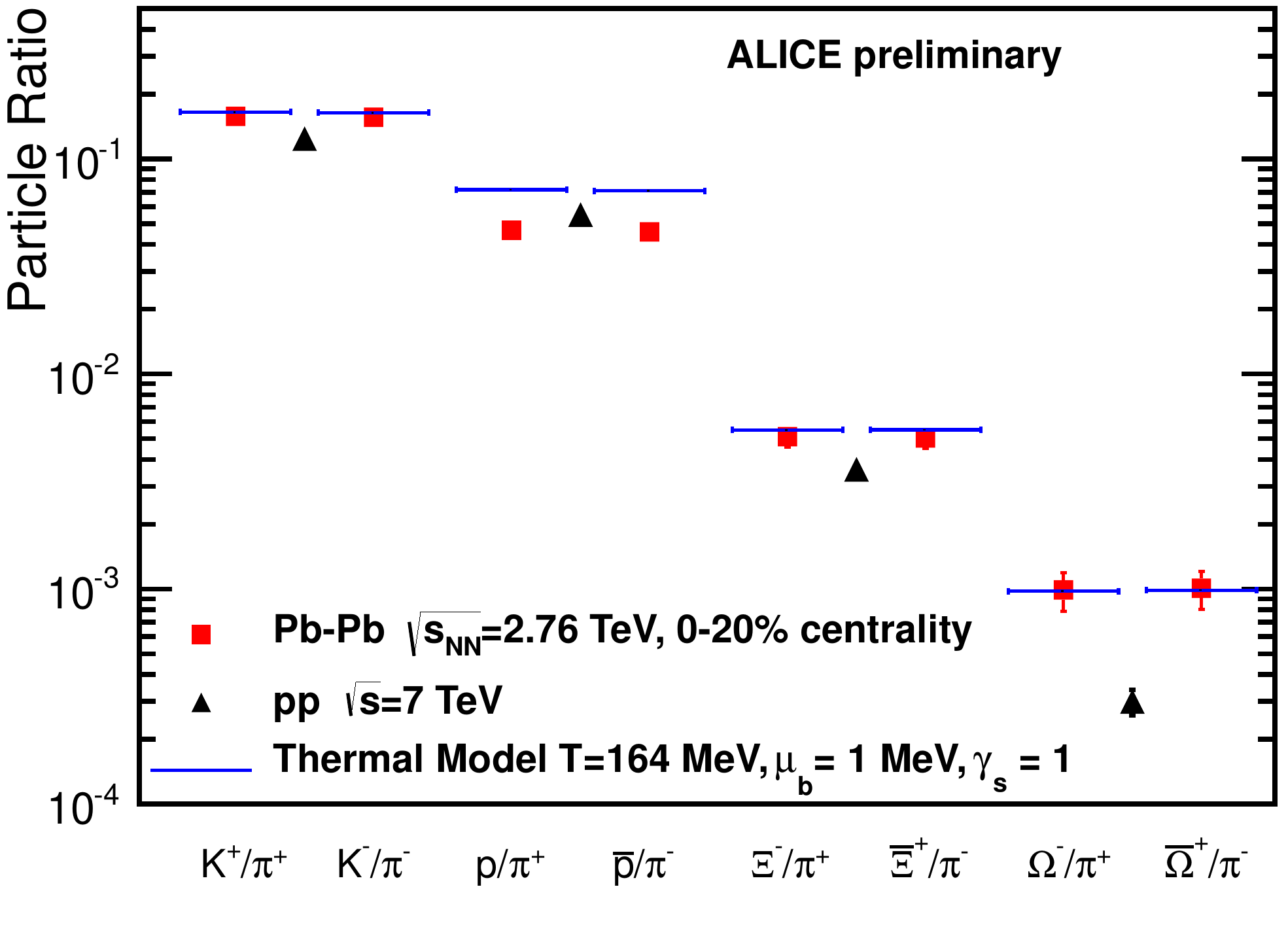}}
\end{minipage}
\end{tabular}

\caption{ a): The \nL/\Kz~ratio as a function of transverse momentum for two centrality selections in nuclear collisions (Pb+Pb at LHC, Au+Au at RHIC) as well as $pp$ collisions at 7 TeV. The full and dashed lines show the corresponding ratios from a hydro and a recombination model (scaled by 0.85), respectively. b) Particle ratios measured in central Pb+Pb (squares) and $pp$ (triangles) collisions at the LHC. The full lines are the predictions from a thermal model.
}
\label{JSPrat}
\end{figure}

The radial flow velocity can be estimated by fitting the spectra with a hydrodynamically inspired function called a ``blastwave'' fit~\cite{Schnedermann:1993ws}. The resulting kinetic freeze-out temperatures $T_{fo}$ and average radial flow velocities $\langle\beta\rangle$ are shown in Fig.~\ref{JSSpect}b for different centrality selections~\cite{Floris:2011ru}. Quantitative results from such fits have to be taken with caution, as they in general depend on the particle types and momentum ranges included in the fit. The two fit parameters ($T_{fo}, \langle\beta\rangle$) are strongly correlated as indicated by the confidence contours in Fig.~\ref{JSPrat}a (each confidence contour corresponds to a different centrality class). However, a relative comparison to a similar fit to RHIC data~\cite{ Adams:2005dq} shows for the most central collisions that the average flow velocity increases significantly at LHC, reaching about $0.65 c$ -- the leading edge of the fireball expands collectively at essentially the speed of light -- and that the kinetic freeze-out temperature drops below the one at RHIC.

Fig.~\ref{JSPrat}a shows the ratio of \nL/\Kz~as a function of \npt~for central and peripheral nucleus-nucleus collisions at LHC and RHIC as well as $pp$ at $\sqrt{s} =7$ TeV~\cite{Belikov:2011zz}. Like the p/\npi~ratio mentioned above, the \nL/\Kz~ratio is strongly enhanced compared to $pp$, reaching a maximum of $\approx 1.5$ around 3 \nGc. Both the maximum value as well as its position in \npt~change only modestly compared to RHIC, where these unusual baryon to meson ratios had been observed previously and were referred to as ``baryon anomaly''~\cite{Adler:2003kg, Abelev:2006jr,Lamont:2007ce}. However, unlike at RHIC, the enhancement for central collisions of roughly a factor of three relative to the $pp$ value persists out to at least 6 \nGc. Qualitatively such an enhancement is expected from radial flow (the heavier baryons are pushed further out in \npt~than the lighter mesons), as shown by the full line, which represents the \nL/\Kz~extracted from the hydro model VISH2+1 already shown in Fig.~\ref{JSSpect}a. Quantitatively, however, the measured shape is very different indicating a progressive decoupling of hadrons from pure hydrodynamic flow above 2 \nGc~ (see Fig.~\ref{JSSpect}a). Global observables like momentum spectra, particle ratios and anisotropic flow  seem to be governed by collective and thermal processes below \npt~$< 2-3$ \nGc, whereas hard processes and fragmentation dominate above \npt~$>6-10$ \nGc. In the intermediate momentum region, various mechanisms have been proposed to describe the interplay between soft and hard processes, including coalescence (recombination) of constituent quarks~\cite{Fries:2008hs}, strong colour fields with enhanced baryon transport~\cite{ToporPop:2011wk}, and colour transparency~\cite{Brodsky:2008qp}. The dashed line shows \nL/\Kz~calculated in a recombination model~\cite{Fries:2008hs} where radial flow and jet-quenching have been adjusted to the measured data at LHC; it describes the shape of the distribution above $\approx$ 3 \nGc~very well but is slightly off in normalisation (the model in the figure has been scaled by 0.85). A quantitative understanding at intermediate momenta remains a challenge and will require additional experimental constraints, in particular high precision data on identified particles out to at least 10 \nGc.

\subsection{Identified particle yields}

Bulk particle production can be very successfully described in the framework of the thermal (statistical) hadronisation model~\cite{BraunMunzinger:2003zd,Becattini:2009sc}. It assumes that particles are created in thermal (phase space) equilibrium governed by a scale parameter T, interpreted as a temperature. Production of a particle with mass $m$ is suppressed by a Boltzmann factor $e^{-m/T}$. Conservation laws introduce additional constraints, like the baryochemical potential $\mu_B$ which accounts for baryon number conservation. An additional parameter $\gamma_s$ is introduced to describe the observation that in some collision systems particles containing strange quarks are suppressed compared to the grand canonical thermal expectation. The temperature parameter is found in all high energy collisions ($pp$, $e^+e^-$, A+A) to be about 160 - 170  MeV, while $\gamma_s$ increases from $0.5-0.6$ in $pp$ to $0.9-1$ in A+A~\cite{Andronic:2009qf,Becattini:2010sk}. The disappearance of strangeness suppression in nuclear collisions~\cite{Blume:2011sb,Margetis:2000sv}, usually called strangeness enhancement, was one of the first signals predicted for the QGP~\cite{Rafelski:1982pu}, and the fact that the bulk of all particles are produced in heavy ion reactions with thermal ratios to very good approximation (typically $<$ 10-20\%) is considered to be an essential and well established fact.

Fig.~\ref{JSPrat}b shows a number of particle ratios from central Pb+Pb collisions, together with the (particle anti-particle averaged) values measured in $pp$ at 7 TeV~\cite{Floris:2011ru,AKSQM2011}, as well as the prediction from a thermal model using the canonical conditions expected for nuclear collisions at the LHC~\cite{Andronic:2008gu}. Within experimental errors, particles and anti-particles are produced in Pb+Pb at midrapidity in equal numbers, consistent with the very small value of the baryochemical potential $\mu_B$ of the thermal model. Strangeness is enhanced in Pb+Pb compared to $pp$, by a factor which increases with strangeness content from about 1.3 in K/\npi~to up to $> 3$ in $\Omega/\pi$ (note that the $pp$ data in Fig.~\ref{JSPrat}b correspond to a higher energy than the Pb+Pb data). All strange particle ratios in Pb+Pb are very well described by the model, implying that they are produced in accordance with fully thermal ratios (i.e. $\gamma_S = 1$). However, in a complete surprise, protons were found to be strikingly off -- too low by a factor of about 1.5 -- and well outside the usual precision of the thermal model. The p/\npi~ratio, which would have been expected to increase from its value measured in $pp$, stays essentially unchanged or even decreases slightly.

Before concluding that something is wrong or missing in current implementations of the thermal model, the data, which are still preliminary, and various corrections (e.g. feed-down from weak decays) have to be thoroughly checked. Also the proton ratios at RHIC should be revisited, as there are indications of a smaller, but still significant, tension between model fits and data~\cite{Aggarwal:2010ig}. If the anomaly persists and cannot be described with extensions to the standard thermal model~\cite{Letessier:2005qe,Rafelski:2010cw}, it may be due to inelastic processes which can change the particle composition after hadronisation. This mechanism is already thought to be responsible for the abundance of resonances with a large width and therefore a large hadronic cross section (like $\rho$ and K$^*$), but hard to describe quantitatively because of partially unknown cross sections and the difficulty to account for multi-particle initial states in hadronic transport codes.

\section{Particle correlations}

\subsection{Identical particle (HBT) correlations}

\begin{figure}[!t]
\begin{tabular}{cc}
\begin{minipage}{.48\textwidth}
\centerline{\includegraphics[width=1.0\textwidth]{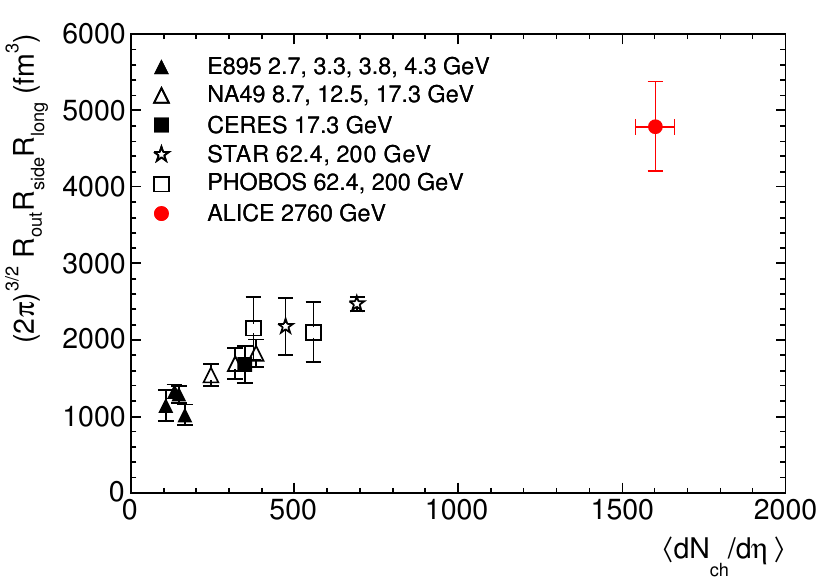}}
\end{minipage} & \begin{minipage}{.48\textwidth}
\centerline{\includegraphics[width=1.0\textwidth]{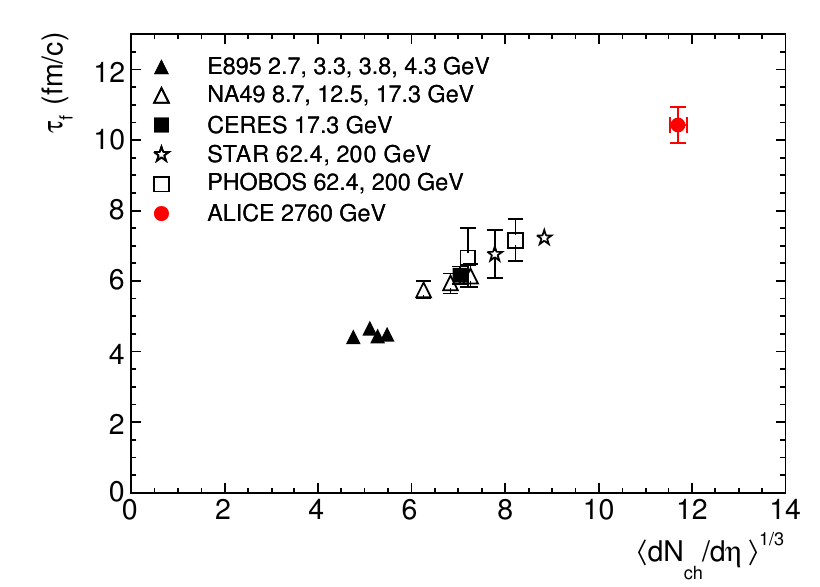}}
\end{minipage}
\end{tabular}

\caption{ a): Local freeze-out volume as measured by identical pion interferometry at LHC compared to central gold and lead collisions at lower energies. b): The system lifetime (decoupling time) $\tau_{f}$ compared to results from lower energies.
}
\label{JSHBT}
\end{figure}

The freeze-out volume (the size of the matter at the time when strong interactions cease) and the total lifetime of the created system (the time between collision and freeze-out) can be measured by identical particle interferometry (also called Hanbury-Brown--Twiss or HBT correlations)~\cite{Lisa:2005dd}. For identical bosons (fermions), quantum statistics leads to an enhancement (depletion) for particles emitted close-by in phase space. This modifies the two-particle correlation function, measured in energy and momentum variables, and can be related via a Fourier transformation to the space and time distribution of the emitting source, i.e. the space-time hyper surface of last rescattering.

Results from HBT correlation measurements are shown in Fig.~\ref{JSHBT} for central collisions from very low energies up to LHC as a function of the charged particle density \dndh~\cite{Aamodt:2011mr}. The total freeze-out volume is given as the product of a geometrical factor and the radii measured in three orthogonal directions (called $R_{\rm long}, R_{\rm side}$, and $R_{\rm out}$), whereas the lifetime was estimated from the pair-momentum dependence of R$_{\rm long}$. The locally comoving freeze-out volume is directly proportional to the particle multiplicity (Fig.~\ref{JSHBT}a) and therefore increases by a factor two compared to top RHIC energy to about 5000 fm$^3$. The system lifetime is proportional to the cube root of the particle density (Fig.~\ref{JSHBT}b) and increases by about 30\% to 10 fm/$c$. Incidentally, the freeze-out volume roughly extrapolates at low beam energy to the volume of a Pb nucleus ($\approx$~800 fm$^3$) and the lifetime vanishes at zero particle density. The evolution from RHIC to LHC of the individual radius parameters ($R_{\rm long}, R_{\rm side}, R_{\rm out}$) as well as their pair momentum dependence is in satisfactory agreement with the predictions of hydrodynamical models~\cite{Abreu:2007kv,Aamodt:2011mr}. The stronger transverse flow generated during the hydrodynamic phase at LHC energy reduces the importance of pre-equilibrium flow, in comparison with RHIC, and thus makes the agreement with the hydrodynamical predictions less dependent on extraneous assumptions \cite{Bozek:2011ph}.

\subsection{Anisotropic Flow}

\subsubsection{General considerations}

The nuclear overlap zone in collisions with non-zero impact parameter is not azimuthally symmetric but has an almond shape whose deformation changes with centrality. Consequently, the pressure gradients between the centre of the overlap zone and its periphery in an average collision vary with azimuth, being strongest in the direction of the reaction plane angle $\Psi_{RP}$, which coincides with the direction of the minor axis of the almond. The developing collective flow is proportional to the pressure gradient and therefore strongest towards the reaction plane, leading to an anisotropic distribution ${\rm d}N/{\rm d}\varphi$ of particles. Anisotropic particle distributions were first suggested in~\cite{Ollitrault:1992bk} as a signal of collective flow in ultra-relativistic heavy ion collisions and the flow pattern is usually quantified via a Fourier expansion~\cite{Voloshin:1994mz}:

\begin{equation}
{\rm E} \frac{{\rm d^3} N}{{\rm d^3}p} =
\frac {1}{2\pi} \frac{{\rm d^2}N}{p_{T} {\rm d}p_{T}{\rm d}y} {\left ( 1 + 2 \sum_{n=1}^{\infty} v_n \cos{[n(\varphi - \Psi_{n})]} \right )}
\label{JSfloweq}
\end{equation}

The Fourier (or flow) coefficients $v_n$ depend on \npt~and are given by
 $v_n(p_T) = \langle \cos{[n(\varphi - \Psi_{n})]}\rangle $,
where the brackets denote an average over particles in a given \npt~bin and over events in a given centrality class. In the above equations, $n$ is the order of the harmonic, $\varphi$  is the azimuthal angle of the particle, and $\Psi_n$ is the angle of the spatial plane of symmetry of harmonic $n$, the plane which maximises the expectation value of $v_n$ in each event. Note that in general the symmetry angles $\Psi_n$ can be independent from each other and point towards different directions in each individual event.

Several experimental methods exist to measure the symmetry plane angles -- using the phi-asymmetry generated by the flow itself to find the directions $\Psi_n$ event-by-event -- and the  coefficients $v_n$  (which are usually averaged over events), via two- or many-particle correlation measures~\cite{Poskanzer:1998yz,Voloshin:2008dg, Snellings:2011sz}.  The $v_1$ coefficient is called {\em directed flow}. It is most prominent near beam rapidity but is not discussed further here (see~\cite{Selyuzhenkov:2011zj}). Most attention was given in the past to $v_2$, the {\em elliptic flow}, which is very strong in non-central collisions and directly linked to the almond shape overlap zone ($\Psi_{2} \approx \Psi_{RP}$). Higher-order harmonics have usually been neglected because they were expected to be small for symmetry reasons (this assumption turned out to be surprisingly poor because of fluctuations, see below).

The elliptic flow magnitude increases continuously with $\sqrt s$ from SPS to RHIC~\cite{Voloshin:2008dg,Snellings:2011sz}. At top RHIC energy, $v_2$ reaches a value compatible with the one predicted by hydrodynamics for a ``perfect fluid'', i.e. a fluid without internal friction and vanishing shear viscosity~\cite{Back:2004je,Arsene:2004fa,Adcox:2004mh,Adams:2005dq,Muller:2006ee}. The shear viscosity, usually quoted in units of $\eta/s$ (shear viscosity $\eta$ over entropy density $s$), for a good relativistic quantum fluid is of order $\hbar/k_B$. Using gauge gravity duality (conformal field theory in Anti de Sitter space, or AdS/CFT) as a proxy for strongly coupled QCD, it has been conjectured that a lower bound exists on $\eta/s$, i.e. $4\pi\eta/s \ge 1 $  ($\hbar = k_B = 1$)~\cite{Kovtun:2004de}.  This value is reached when the 't Hooft coupling  tends to infinity, and the mean free path approaches the quantum limit, the Compton wavelength. In such a ``perfect'' fluid, pressure gradients are not dissipated away because momentum transport perpendicular to the gradient is restricted by the ultra-short mean free path. 

Bounds on the $\eta/s$ ratio can be derived from the data in two different ways. Either one compares the momentum dependent elliptic flow parameter $v_2(p_T)$ with results from viscous hydrodynamics calculations, or one extracts a value of $\eta/s$ by fitting the centrality (i.~e.\ transverse size) dependence of the average $p_T$ integrated elliptic flow. Both approaches yield a $\eta/s$ ratio at RHIC that is very close to the lower bound, at most 3-5 times $1/4\pi$~\cite{Teaney:2009qa,Nagle:2009ip}. The most advanced model--data comparison using the second method gives an even tighter bound: $4\pi\eta/s \leq 2.5$ \cite{Song:2010mg}.

\begin{figure}[!t]
\begin{tabular}{cc}
\begin{minipage}{.48\textwidth}
\centerline{\includegraphics[width=1.0\textwidth]{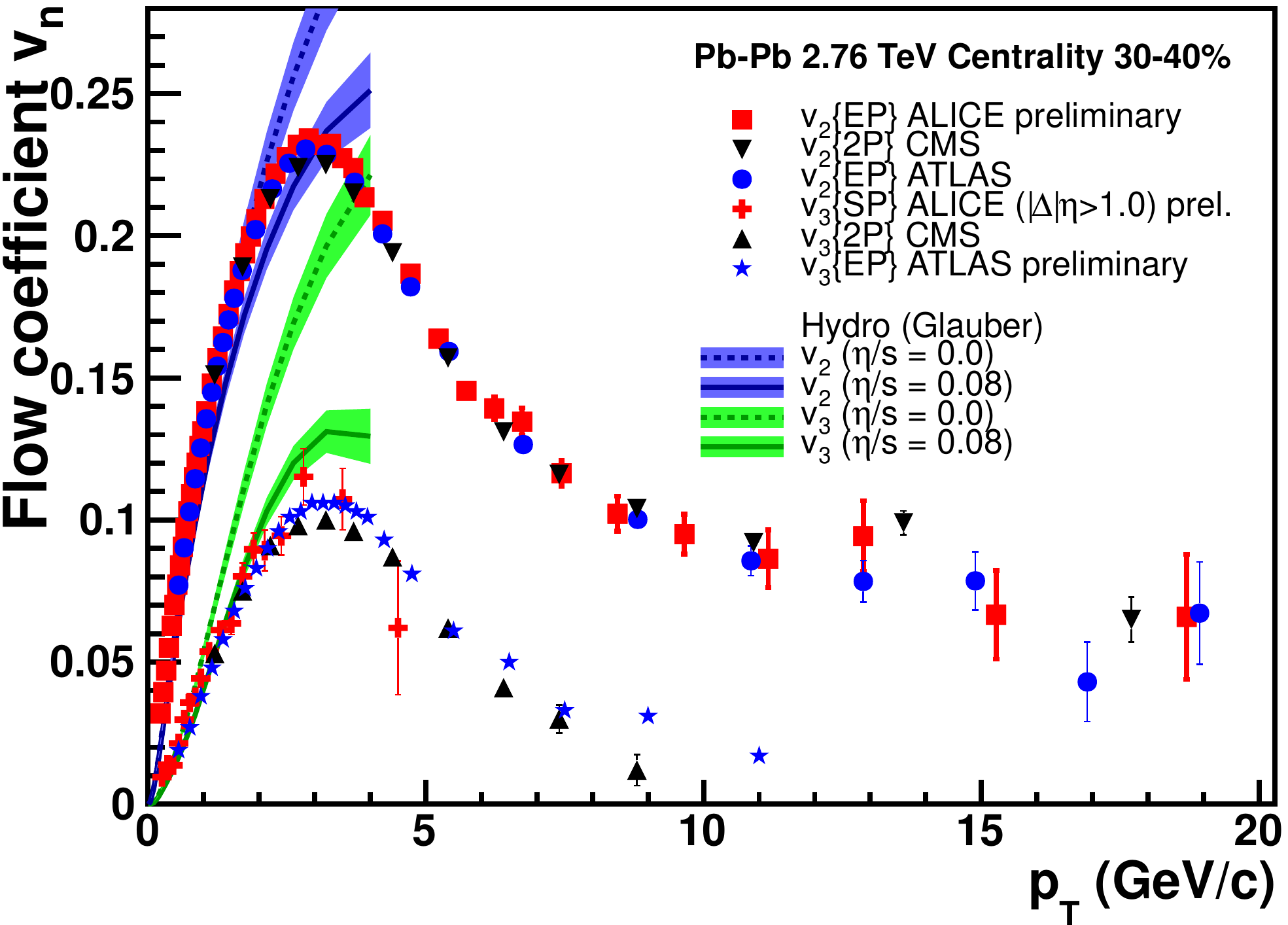}}
\end{minipage} & \begin{minipage}{.48\textwidth}
\centerline{\includegraphics[width=1.0\textwidth]{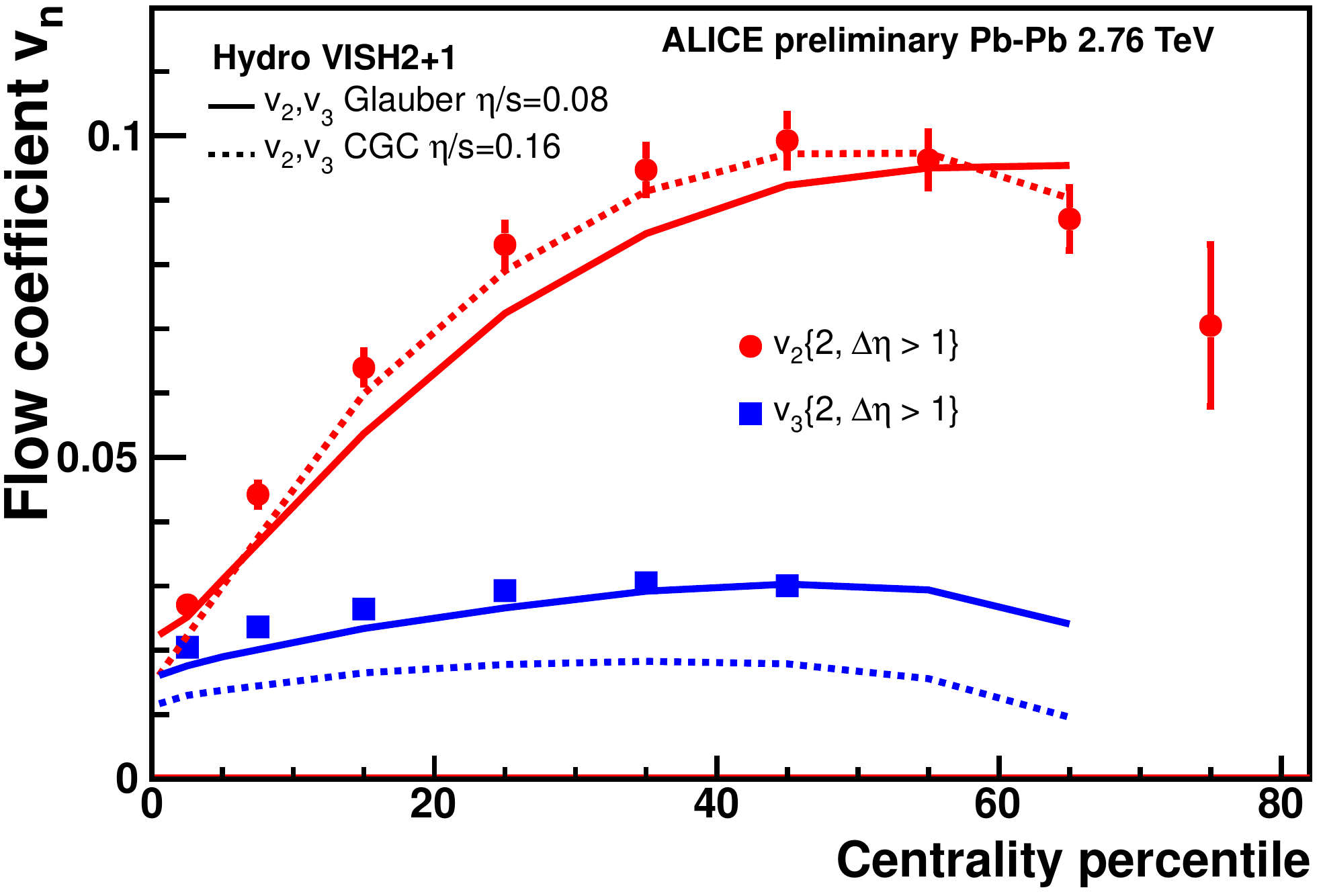}}
\end{minipage}
\end{tabular}

\caption{ a): Elliptic flow coefficients $v_2$ and $v_3$ as a function of transverse momentum compared to a hydro model~\cite{Schenke:2011tv} using two different values for the viscosity $\eta/s$ b): Integrated elliptic flow coefficients as a function of collision centrality compared to a hydro calculation~\cite{Qiu:2011hf} using two different models for the initial state geometry.
}
\label{JSflow1}
\end{figure}

Assuming no or only small changes in the transport properties of the quark-gluon plasma between RHIC and LHC, hydrodynamic models predicted that the (particle type averaged) elliptic flow coefficient $v_2$, measured as a function of $p_T$, should change very little, whereas the momentum integrated flow values should either stay approximately constant or rise by at most 30\%~\cite{Abreu:2007kv}. This prediction was quickly confirmed~\cite{Aamodt:2010pa}: the $p_T$ differential flow for charged particles is essentially unchanged at LHC whereas the $p_T$ integrated flow  increases by some 30\%, a direct consequence of the stronger radial flow discussed above. The matter created at LHC therefore still behaves like the (almost) perfect liquid discovered at RHIC. 

In order to make progress in the precision determination of $\eta/s$, some major obstacles have to be overcome.  An example on the experimental side are non-flow correlations introduced e.g. by jets and resonance decays.  On the theory side, uncertainties arise from model assumptions made about the initial state pressure gradients and their event-by-event fluctuations. It was in fact only recently realized~\cite{Sorensen:2010zq,Alver:2010gr} that the statistical nature of individual nucleon-nucleon collisions can lead to highly irregular shapes of the reaction zone and the corresponding initial  energy/pressure distributions. These shapes fluctuate from one event to the next, even at a fixed impact parameter. The irregular pressure gradients show no symmetry with respect to the reaction plane and therefore induce higher harmonic flow patterns. Fluctuations also tilt the symmetry angle of $v_2$ away from the geometrical reaction plane ($\Psi_2 \ne \Psi_{RP}$), leading to event-by-event fluctuations of the elliptic flow direction and magnitude.

\subsubsection{Charged particle anisotropic flow}

The elliptic ($v_2$) and triangular ($v_3$) flow coefficients for mid-central Pb-Pb collisions are shown in Fig.~\ref{JSflow1}a as functions of \npt.  The agreement between the LHC experiments~\cite{ALICE:2011ab,ADQM2011,ATLAS:2011ah,Trzupek:2011zz,ATLAS-CONF-2011-074,Chatrchyan:2012wg} is again remarkable. Making use of the large rapidity acceptance of the LHC detectors, the coefficients $v_n$ were measured by methods that minimize the influence of non-flow correlations, which are dominantly short-range in rapidity (2P = trigger di-hardon correlations, EP = Event Plane, SP = two-particle cumulant with eta-gap; see~\cite{Voloshin:2008dg}). The elliptic flow coefficient rises approximately linearly with \npt~to a maximum of $v_2 \approx 0.23$ around 3 \nGc. The corresponding asymmetry in ${\rm d}N/{\rm d}\varphi$ is very large indeed: almost three times as many 3 \nGc~particles are emitted in-plane compared to out-of-plane (see eq~\ref{JSfloweq}). The coefficient then decreases, at first rapidly then more gradually, but stays finite out to the highest \npt~measured. Also the triangular flow coefficient is very significant out to about 10 \nGc~ and similar in shape, reaching about half the value of $v_2$ at the maximum.

At high $p_T$, where spectral shapes and relative particle yields indicate that hadrons are no longer in local thermal equilibrium, the nonzero $v_2$ is thought to arise from differential parton energy loss rather than collective flow. The energy loss should depend on the path length inside the matter and therefore correlate naturally with the orientation of the elongated reaction zone.

At lower $p_T$, the data are compared to a hydrodynamic calculation~\cite{Schenke:2011tv} using two different values of the shear viscosity($4\pi\eta/s = 0$ and $1$). The effect of a nonvanishing viscosity is clearly visible: The initial pressure gradients are dissipated and the resulting collective flow is reduced. The influence of viscosity is seen to be much stronger on $v_3$, as expected. The triangularity corresponds to shorter wavelength density variations and therefore larger local pressure gradients, which are more sensitive to viscosity.  While both values of the viscosity are compatible with the $v_2$ results, only $4\pi\eta/s = 1$ describes $v_3$.  A larger value of the viscosity would fall significantly below the $v_3$ data, but could in principle be accommodated by using a different model for the initial state which generates a larger pressure gradient. 

This is illustrated in Fig.~\ref{JSflow1}b, which shows the $p_T$ integrated flow as a function of centrality~\cite{ALICE:2011ab}. The elliptic flow rises strongly towards peripheral collisions, following the increasing elongation of the reaction region.  Because of event-by-event fluctuations $v_2$ stays non-zero even for the most central collisions.  The triangular flow shows little centrality dependence, as it is driven solely by shape fluctuations. The hydrodynamic model~\cite{Qiu:2011hf} shown in Fig.~\ref{JSflow1}b employs two different, frequently used sets of initial conditions. The energy density distribution in the transverse plane is taken either from a Glauber calculation (solid line) combined with $4\pi\eta/s = 1$, or from a saturation model (dashed line) combined with $4\pi\eta/s = 2$. The latter predicts larger pressure gradients and consequently needs a larger viscosity to describe the $v_2$ data. However, the saturation model significantly underpredicts $v_3$. Glauber initial conditions can describe both $v_2$ and $v_3$ better, in particular, if flow fluctuations are taken into account (see~\cite{Qiu:2011hf} for details). 

The measurement of higher harmonics thus seems to be able to overcome the degeneracy between shear viscosity and initial conditions and give a tighter limit of $4\pi\eta/s \le 2$~\cite{Qiu:2011hf}. Recent measurements of higher harmonic flow at RHIC have led to a similar conclusion~\cite{Adare:2011tg,Sorensen:2011fb}.  However, none of the currently used initial state models can perfectly describe all experimental flow observations. With the on-going effort in both experiment and theory, the new high precision flow data from LHC promise significant further progress toward the goal of a precision measurement of the viscosity and other properties of the quark-gluon plasma.

\begin{figure}[!t]
\begin{tabular}{cc}
\begin{minipage}{.48\textwidth}
\centerline{\includegraphics[width=1.0\textwidth]{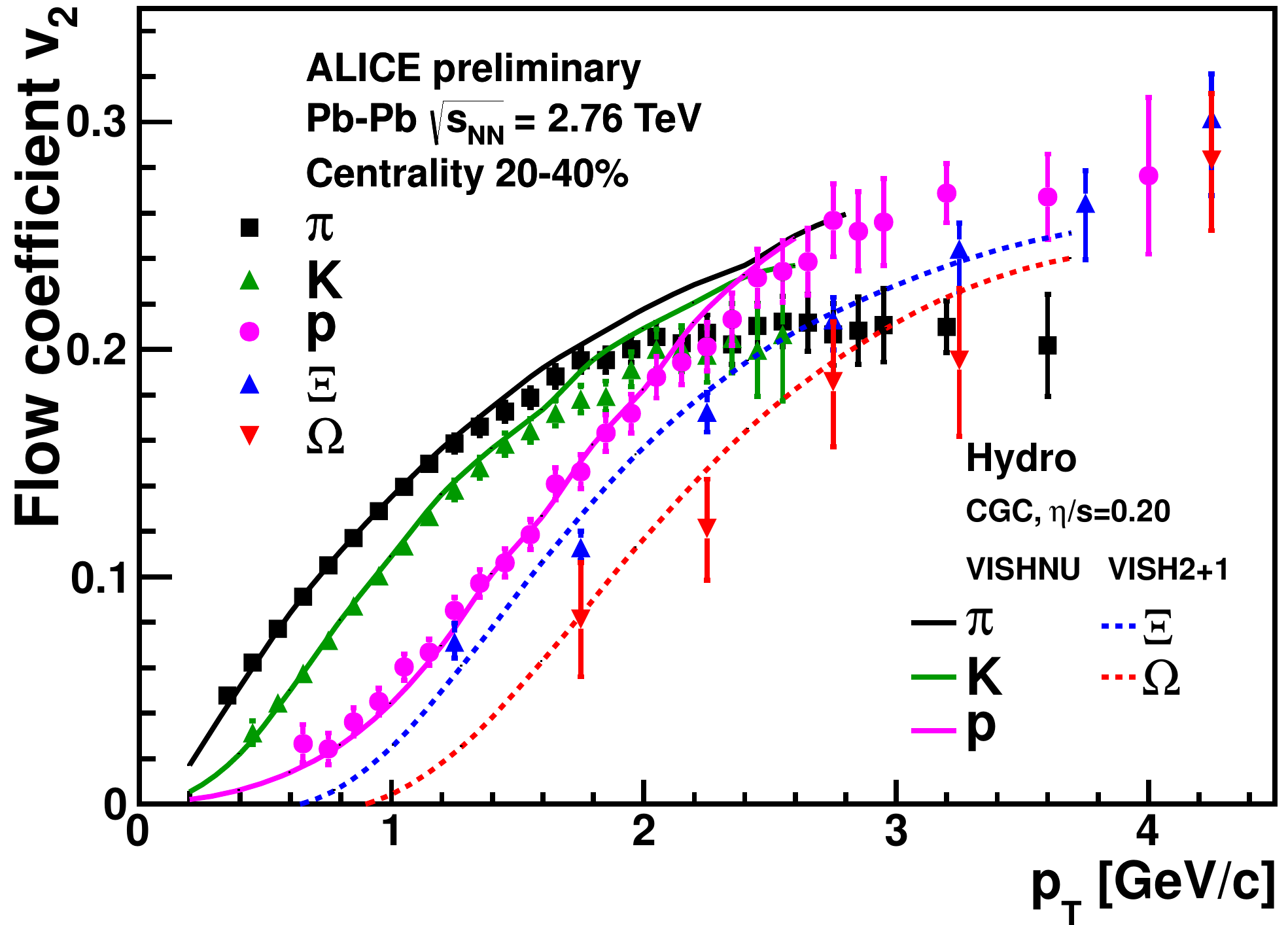}}
\end{minipage} & \begin{minipage}{.48\textwidth}
\centerline{\includegraphics[width=1.0\textwidth]{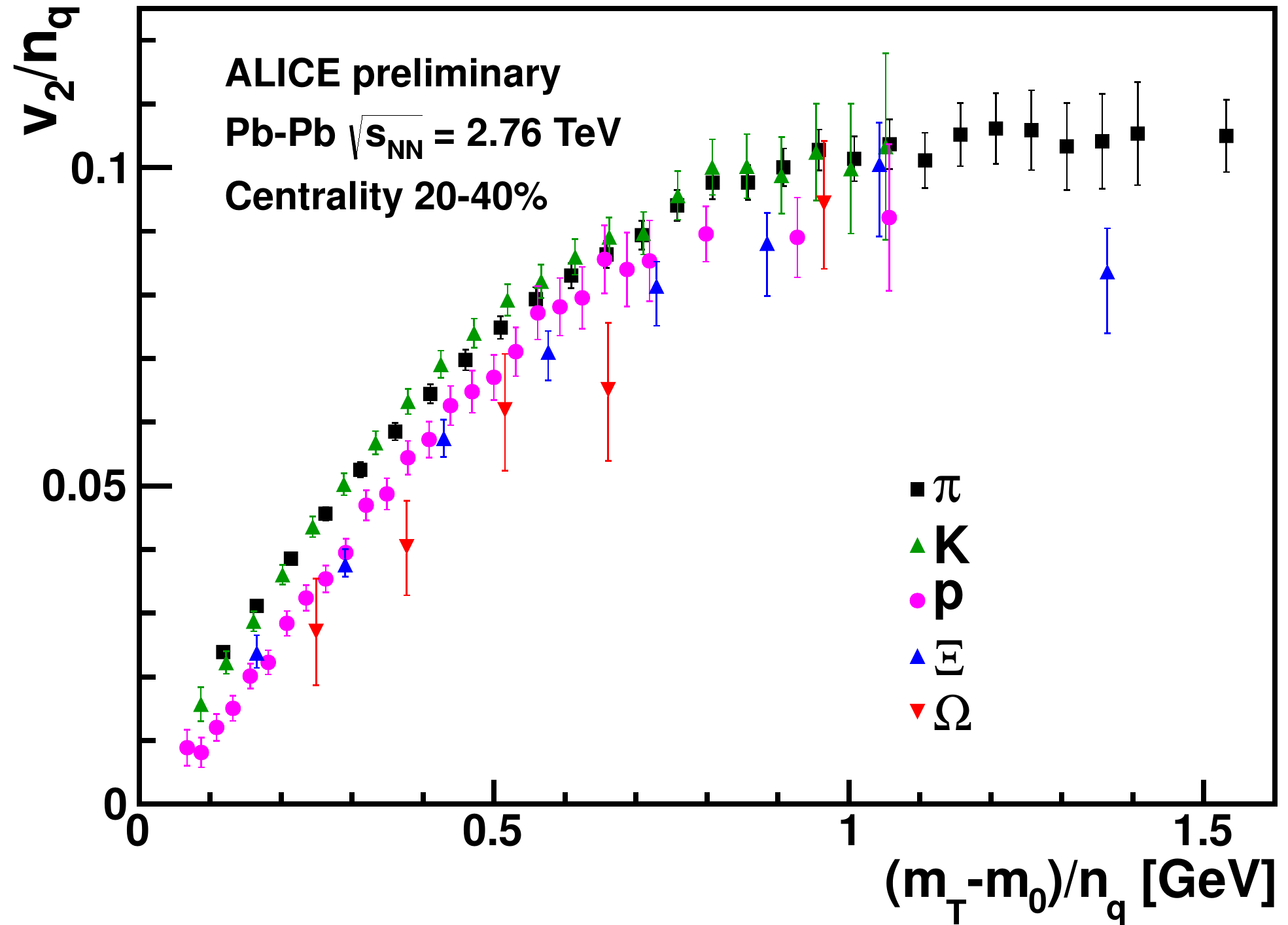}}
\end{minipage}
\end{tabular}
\caption{ 
a): Elliptic flow $v_2$ as a function of \npt~for identified particles compared with a hydrodynamical model~\cite{Shen:2011eg, Heinz:2011kt}.
b): Elliptic flow for mesons and baryons, scaled by the respective number of valence quarks $n_q$, versus scaled transverse kinetic energy.
}
\label{JSflow2}
\end{figure}

The interference between $v_2$ and higher harmonic flow coefficients~\cite{ALICE:2011ab,Jia:2011hfa,Chatrchyan:2012wg,Adare:2011hd,Li:2011mp}, which are significant up to $n=6$, also gives a most natural explanation for some unusual long range structures observed in two-particle correlation at RHIC~\cite{Sorensen:2010zq}, traditionally called the ``soft near-side ridge'' and the ``away-side cone''.  The interpretation of these structures was controversial \cite{Nagle:2009wr} until higher harmonic flow was firmly established at LHC and simultaneously at RHIC.

\subsubsection{ Identified particle flow}

As for radial flow, a most stringent test of the collective flow interpretation of azimuthal anisotropies is the characteristic dependence on particle mass. The LHC data for $v_2$ for various particles (\npi, K, p, $\Xi, \Omega$) are shown in  Fig.~\ref{JSflow2}a~\cite{Krzewicki:2011ee,Yin:2012sk}.  To understand the origin of the characteristic mass splitting, seen also in $v_3$~\cite{Krzewicki:2011ee}, one needs to keep in mind that collective radial flow tends to equalise the  velocities of particles, not their momenta, and therefore shifts heavy particles out to higher \npt~than light ones~\cite{Huovinen:2001cy}. The effect of a given azimuthal flow asymmetry thus manifests itself at higher momenta for particles with a larger mass. The hydrodynamical model~\cite{Heinz:2011kt, Shen:2011eg}, which incorporates this effect, describes the data very well for all particle species up to intermediate \npt.  It also predicted the observation that the mass splitting is larger at LHC than at RHIC as a consequence of the increased radial flow. 

The mesons (\npi, K) deviate from the predicted flow above 1.5 \nGc, whereas the baryons follow the hydrodynamical curves, within the still large experimental errors, out to about 3 \nGc. The different behaviour of mesons and baryons, also seen in RHIC data~\cite{Back:2004je,Arsene:2004fa,Adcox:2004mh,Adams:2005dq,Muller:2006ee}, has been interpreted as a sign of quark recombination or coalescence \cite{ Voloshin:2002wa,Fries:2003vb,Fries:2003kq,Greco:2003mm}, already mentioned above as a possible reason for the large baryon-to-meson ratios. In this picture, hadrons at intermediate \npt~of a few \nGc~are formed by coalescence of constituent quarks from deconfined quark matter, with the three-quark baryons receiving a proportionally larger share of the collective quark flow than the two-quark mesons. Indeed, RHIC data indicate a universal scaling of flow \cite{Adare:2006ti} for all particles of mass $m$ when plotting $v_2/n_q$ versus $(m_T-m)/n_q)$, where $n_q$ is the number of constituent quarks ($n_q$~= 2 or 3) and $(m_T - m)$ is the transverse kinetic energy ($m_T = \sqrt{p_T^2 + m^2}$). 

The same scaled distribution is shown in Fig.~\ref{JSflow2}b~\cite{Krzewicki:2011ee,Yin:2012sk} for LHC data. No scaling is apparent for $p_T < 1.5$ \nGc, where all particles follow more closely the mass dependent pattern predicted by hydrodynamics (see Fig.~\ref{JSflow2}a). It therefore seems that the empirically observed low \npt~ ``quark scaling'' at RHIC is a coincidence, only approximately valid for a particular value of the radial flow and the corresponding mass splitting.  At higher \npt, quark scaling between mesons and baryons may hold at LHC within the current experimental errors, but is definitely less convincing so far. Reduced errors and a measurement of flow for the $\phi$, a meson with a mass comparable to the proton, are required before a strong case for the coalescence picture can be made with LHC flow results.

\section{Hard processes at high $p_T$ and high mass}

The increased energy of heavy ion collisions at LHC relative to RHIC leads to much larger cross sections for hard processes, i.e. those involving high momentum or high mass scales. Energetic quarks or gluons can be observed as jets or single particles with $p_T$ reaching 100 \nGc~and beyond. Similarly, high $p_T$ photons, charmonium and bottonium states (i.e. the \Jpsi~and \nY~families), and even the weak vector bosons $W$ and $Z$ are copiously produced. The details of production and propagation of these high $p_T$ probes can be used to explore the mechanisms of parton energy loss and deconfinement in the medium. 

Some of the salient questions to be explored in this way are: How do colour charge and mass of a parton influence the energy loss? The perturbative QCD formalism of radiative energy loss \cite{Majumder:2010qh} predicts that the energy loss of a parton should be proportional to the Casimir eigenvalue of its colour charge, i.~e.\ that gluons should lose energy at roughly twice the rate of quarks, and that the rate of energy loss of heavy quarks should be reduced by the so-called dead-cone effect \cite{Dokshitzer:1991fd}. What is the relative contribution of radiative and elastic mechanisms of energy loss?  How does the energy loss depend on the thickness of the medium? Strongly coupled gauge theories generally predict a stronger dependence on the path length in the medium than perturbative QCD \cite{Muller:2010pm}. Where does the lost energy go? Does it remain inside the jet cone and manifest itself as a modification of the fragmentation function, or does it get rapidly thermalized in the medium, making the jet look like a jet in vacuum, but with reduced energy?

There are several experimental techniques available to explore these questions at the LHC. One is to measure single particle cross sections and compare them to the equivalent cross sections in $pp$ collisions. The nuclear modification factor $R_{AA}$ of single particles, which has been extensively studied at RHIC~\cite{Back:2004je,Arsene:2004fa,Adcox:2004mh,Adams:2005dq}, allows for direct comparison with the RHIC results and their extension to higher momenta.  The other approach is to reconstruct jets directly and compare jet yields in Pb+Pb with jet production in $pp$, as well as to measure jet--jet and jet--charged particle energy and angular correlations. The large kinematic range and the excellent calorimetry of the LHC detectors facilitate jet reconstruction, especially for jet energies in excess of 50 GeV.

\subsection{Single particle spectra}

The single particle production rates at RHIC have shown a large suppression of hadrons in nuclear collisions relative to $pp$, whereas particles that do not interact strongly, e.~g.\ photons, are not modified. The LHC can significantly extend the accessible $p_T$ range and allow the measurement of additional particles, such as the $Z$ and $W$. The suppression effects of a given particle are typically expressed in terms of the nuclear modification ratio:
\begin{equation}
{R_{AA}(p_T)}={{d^{2}N_{AA}/d{p_T}d{\eta}}
\over{\langle T_{AA}\, \rangle d^{2}\sigma_{NN}/d{p_T}d{\eta}}} ,
\end{equation}
where $N_{AA}$ and $\sigma_{NN}$ represent the particle yield in nucleus-nucleus collisions and the cross section in nucleon-nucleon collisions, respectively. The nuclear overlap function $\langle T_{AA} \rangle$ is the ratio of the number of binary nucleon-nucleon collisions, $\langle N_{\rm coll} \rangle$, calculated from the Glauber model, and the inelastic nucleon-nucleon cross section ($\sigma^{NN}_{\rm inel}=(64\pm 5)$ mb at $\sqrt{s_{NN}}={\rm 2.76}$ TeV). In the absence of nuclear effects the factor $R_{AA}$ is unity by construction. As observed at RHIC in 200 GeV Au+Au collisions~\cite{ Adcox:2004mh,Adams:2005dq}, the yield of $5-10$ GeV/c charged particles is suppressed in the most central events by more than a factor of five. Instead of $R_{AA}$ one can also approximate the centrality dependence by measuring $R_{CP}$, the ratio of central over peripheral events.

While the charged particle $R_{AA}$ is the measurement with best statistical and systematic precision~\cite{Aamodt:2010jd, Appelshauser:2011ds, Collaboration:2012nt, Milov:2011jk}, it is also interesting to measure the nuclear suppression factor for individual particle species to distinguish the exact mechanisms of energy loss. Measurements exist for identified $\pi$, \Kz, \nL~\cite{Appelshauser:2011ds,Dainese:2011vb}, isolated photons~\cite{Chatrchyan:2012vq}, $Z, W$~\cite{ Steinberg:2011dj,Chatrchyan:2011ua,ATLAS:2010px,ATLAS-CONF-2011-078}, $D$-mesons~\cite{Dainese:2011vb}, jets~\cite{Milov:2011jk}, $J/\psi$~\cite{Abelev:2012rv,ATLAS:2010px, Chatrchyan:2012np} and $\Upsilon$~\cite{Chatrchyan:2012np}. The suppression factors of prompt and non-prompt $J/\psi$ mesons, the latter being produced from $B$-mesons and identified by their displaced decay vertex, were measured separately. In addition, ATLAS has presented results for the relative suppression of charged particles in central versus peripheral events, $R_{CP}$ \cite{ATLAS-CONF-2011-079}.

\begin{figure}[!t]
\begin{tabular}{cc}
\begin{minipage}{.50\textwidth}
\centerline{\includegraphics[width=1.0\textwidth]{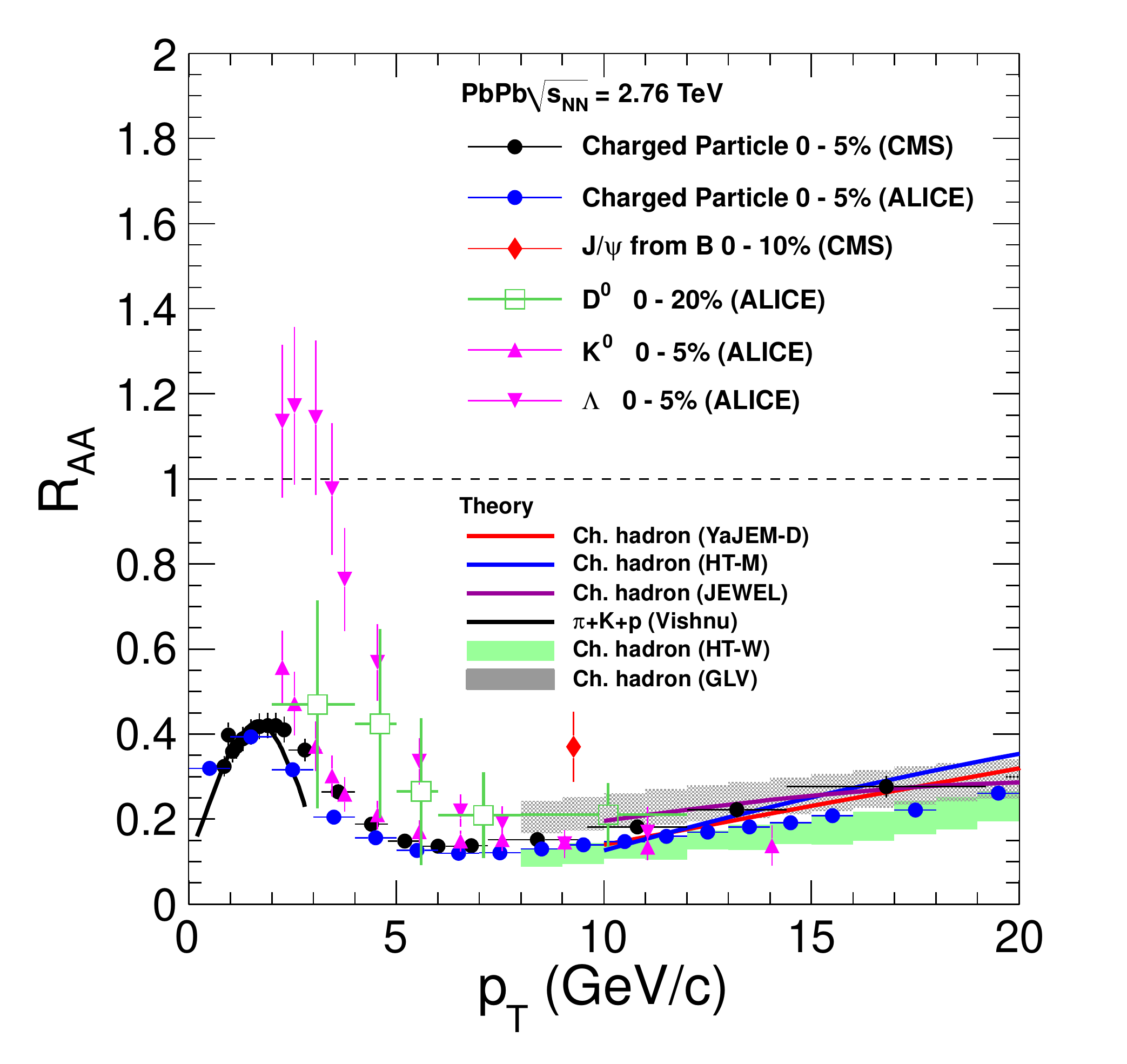}}
\end{minipage} & \begin{minipage}{.50\textwidth}
\centerline{\includegraphics[width=1.0\textwidth]{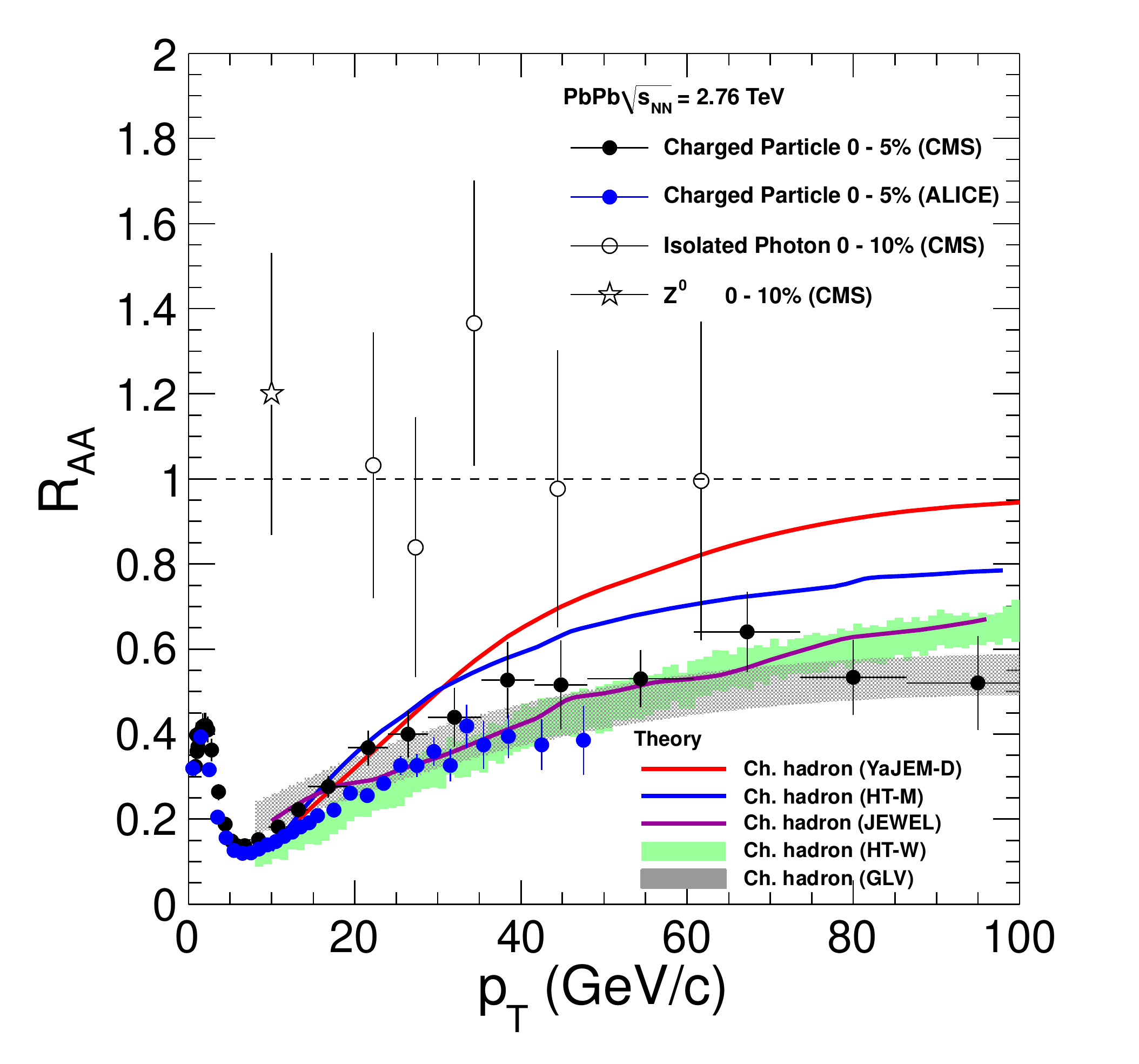}}
\end{minipage}
\end{tabular}
      	\caption{Nuclear modification factor $\RAA$ as a function of $p_T$ for a variety of particle species together with theoretical predictions. Experimental error bars correspond to the total error (statistical  and systematic errors added in quadrature).
	a) Low momentum region $p_T < 20$ GeV; b) Entire momentum range measured at LHC. The curves show the results of various QCD-based models of parton energy loss \cite{Sharma:2009hn,Renk:2011gj,Chen:2011vt,Majumder:2011uk,Zapp:2011ek}. For details, see text.}
\label{fig:raa}
\end{figure}

A summary of $R_{AA}$ measurements  for different particle species is shown in Fig.~\ref{fig:raa} for the most central events. The inclusive charged particle $R_{AA}$ follows, up to about $10-15$ \nGc, the characteristic shape discovered at RHIC (left panel, full circles). The pronounced maximum at a few \nGc, which is sometimes attributed to initial or final state interactions in nuclei (``Cronin effect''), is at very high energies more likely to be yet another manifestation of collective flow. It is qualitatively described by the dashed line, which shows the $R_{AA}$ obtained by dividing the inclusive charged particle distribution calculated by viscous hydrodynamics \cite{Song:2007ux} for central Pb+Pb by the experimentally measured $pp$ spectrum.  This interpretation is also supported by the fact that the apparent ``suppression'' factor is slightly larger for kaons and significantly larger for the \nL, as expected from flow. The peak region is followed by a steep decline and a minimum, around $5-7$ \nGc, where the suppression reaches a factor of about seven, very similar to but slightly larger than the one measured at RHIC.

Heavy quarks, as shown by the $R_{AA}$ of prompt D mesons (open squares) and non-prompt \Jpsi~(from the decay of bottom quarks, closed diamond) in Fig.~\ref{fig:raa}, are almost as strongly suppressed as inclusive charged particles. A similar conclusion can be drawn from the measurement of leptons from heavy flavour decays~\cite{Dainese:2011vb}. This seems contrary to the expectation that gluons, which are the dominant source of inclusive charged particles at LHC, should suffer twice as much energy loss as light quarks and that, in addition, the energy loss of heavy quarks should be even less than that of light quarks because of the mass dependence of radiation (``dead-cone'' effect \cite{Dokshitzer:1991fd}). The strong suppression found for hadrons containing $c$- and $b$-quarks confirms observations made at RHIC and may indicate that the energy loss rate depends less strongly on the parton mass than expected for radiative energy loss. Reasons for this behaviour could be nonperturbatively large elastic energy loss in the strongly coupled quark-gluon plasma or heavy meson formation within the medium~\cite{Sharma:2009hn}. More data and a quantitative comparison with models will be required to see how the small, with current statistics not very significant, difference between light hadron and heavy quark suppression can be accommodated by theory.

Above $p_T \approx 8$~GeV/$c$, the suppression becomes universal for all particle species (with the possible exception of the non-prompt \Jpsi originating from $B$-meson decays shown in the left panel). With increasing $p_T$, $R_{AA}$ rises gradually towards a value of 0.5 (see right panel), a feature which was not readily apparent in the RHIC data. Isolated photons and the $Z$ boson are not suppressed, within the currently still large statistical errors. This finding is consistent with the hypothesis that the suppression observed for hadrons is due to final-state interactions with the hot medium.  

The observed rise of $R_{AA}$ with $p_T$  allows a better discrimination between competing models of energy loss than the rather flat high $p_T$ dependence observed at RHIC. The rise can be understood as a decrease of the parton fractional energy loss with increasing $p_T$, reflecting the weak energy dependence of pQCD radiative energy loss on parton energy.  At RHIC this trend is compensated by the softening of the underlying parton spectrum, whereas at LHC the spectrum stays hard up to the highest measured $p_T$ which remains much farther away from the kinematic threshold than at RHIC. 

The observed trend is semi-quantitatively described by several models implementing the perturbative QCD (pQCD) formalism for energy loss~\cite{Sharma:2009hn,Renk:2011gj,Chen:2011vt,Majumder:2011uk,Zapp:2011ek}. The rate of induced gluon radiation in pQCD is governed by the rate of transverse momentum broadening, encoded in the jet quenching parameter $\hat{q}$, of the trajectory of a hard parton in the quark-gluon plasma \cite{Majumder:2010qh}.  The value of $\hat{q}$ is proportional to the density of gluons in the medium and grows approximately as the entropy density of the plasma, or the charged particle multiplicity in the final state of the heavy ion collision. It should thus be slightly more than twice as large in Pb+Pb collisions at LHC than at the top RHIC energy.  

The band of curves labelled ``GLV'' \cite{Sharma:2009hn} in Figure~\ref{fig:raa} shows results of a pQCD calculation using light-cone wave functions for the produced hadrons and assuming static scattering centres corresponding to a gluon density $dN_g/dy \approx 2,800$ and a QCD coupling constant $\alpha_s \approx 0.3$. The curve labeled ``YaJEM-D'' \cite{Renk:2011gj} is a Monte-Carlo implementation of pQCD jet quenching that accounts for the limited virtuality evolution of the jet in a finite medium.  The curves labelled ``HT-W'' \cite{Chen:2011vt} are based on the higher-twist formalism and correspond to values of $\hat{q}_0\tau_0 = 1.0 - 1.4~{\rm GeV}^2$ at the time of thermalization $\tau_0$. The curve labelled ``HT-M'' \cite{Majumder:2011uk} is a different implementation of the higher-twist formalism with the parameter choice $\hat{q}_0 = 1.3~{\rm GeV}^2$/fm at $T_0=344$ MeV. The ``JEWEL'' model \cite{Zapp:2011ek} uses a  Monte-Carlo implementation of an in-medium parton shower including LPM radiation suppression.

The various calculations apply models of the nuclear reaction that are of widely varying detail. The GLV model uses a schematic description of the integrated medium density based on the boost-invariant Bjorken model of hydrodynamics. The YaJEM-D model combines ideal two-dimensional hydrodynamics with a saturation model for the initial energy density, but has not adjusted the parameters to the measured global properties of Pb+Pb collisions at LHC. The HT-W and HT-M calculations use three-dimensional ideal and two-dimensional viscous hydrodynamics, respectively, with initial conditions chosen to reproduce the charged particle multiplicity measured at LHC.  The JEWEL model uses a variant of the schematic Bjorken hydrodynamical model.

It is an important theoretical challenge to identify the origin of the differences between the five calculations, in particular, the much more rapid rise of $R_{AA}$ in the YaJEM-D and HT-M models. Systematic studies of the differences between various implementations of radiative parton energy loss in pQCD at RHIC energies have revealed large sensitivities to poorly controlled aspects of lowest-order, collinear gluon radiation. These differences can be further enhanced by the use of different models for the medium evolution. Nevertheless, the quantitative success of several models in correctly reproducing the rise and saturation of $R_{AA}$ with $p_T$ suggests that the energy loss of the leading parton in a jet shower may be described by perturbative QCD radiation in a strongly coupled medium.

\subsection{Jets}

Studying the modification of fully reconstructed jets is a particularly useful tool for probing the properties of the hot quark-gluon plasma. Jets are formed by fragmentation from high $\PT$ partons as they propagate through the produced matter.   Measuring the energy of fully reconstructed jets allows one to distinguish between energy redistribution among the leading parton and the remainder of the jet and energy dissipation out of the jet into the thermal medium.  One of the promising channels are dijets, in particular their transverse energy balance and azimuthal angle correlation. The energy dissipation into the medium can be studied by measuring the asymmetry in $\PT$ of dijets in heavy ion collisions as a function of centrality and by comparing them to data from $pp$  collisions. Similarly, the measurement of the relative azimuthal angle of the two jet axes gives information about the degree of scattering of partons as they traverse the medium.

The measurement of the dijet asymmetry $A_J= (p_{T1}-p_{T2})/(p_{T1}+p_{T2})$, where ``1'' and ``2'' refer to the leading and subleading jet, respectively, was performed by both ATLAS \cite{Aad:2010bu} and CMS \cite{Chatrchyan:2011sx}.  Events containing at least two jets, with the leading (sub-leading) jet having $\PT$ of at least 120 (50) \nGc~for CMS and at least 100 (25) \nGc~for ATLAS, were selected for further study.

\begin{figure}[ht!]
\begin{center}
   	\includegraphics[width=0.9\textwidth]{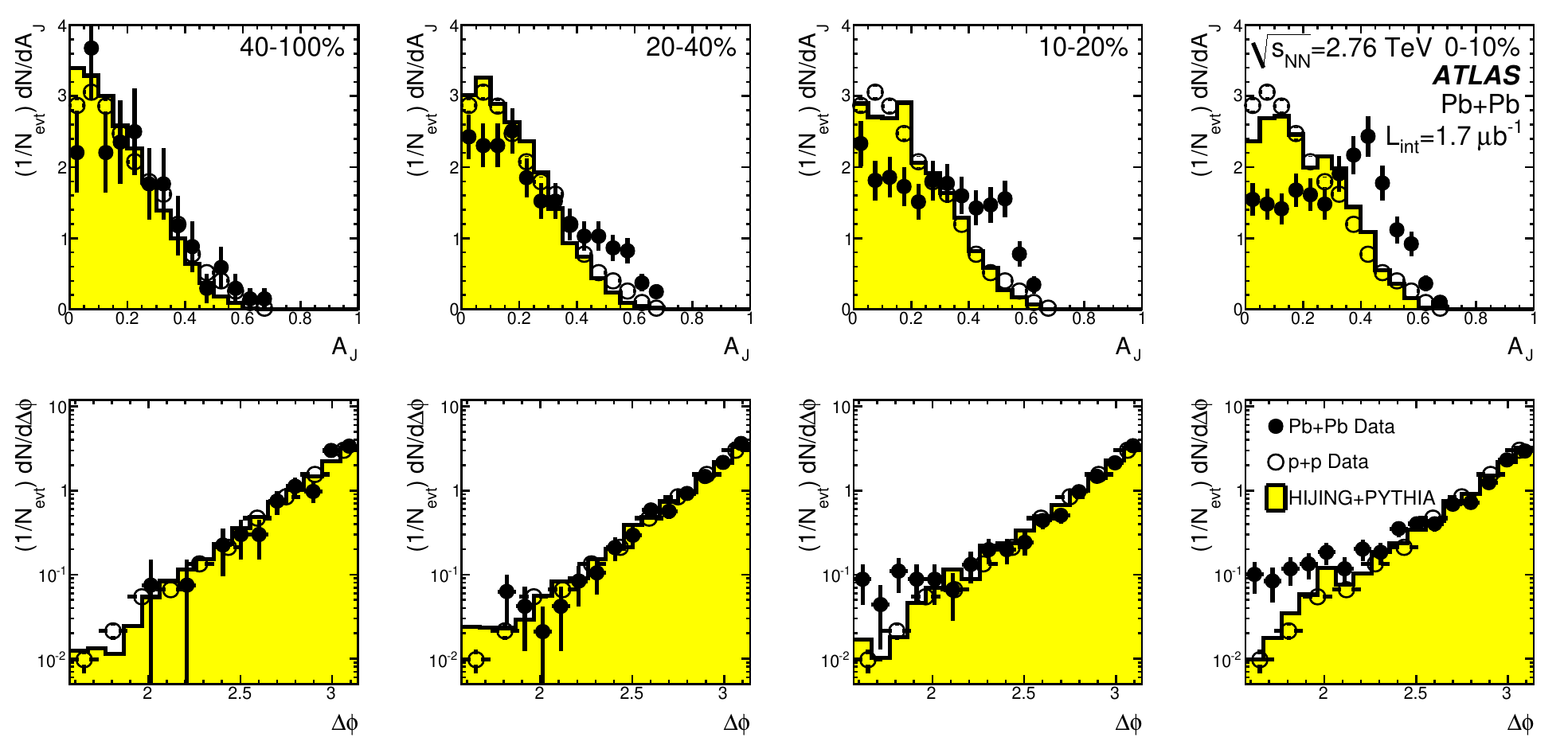}
       \caption{Calorimetric jet imbalance in dijet events (top) and azimuthal angle between the leading and subleading jets (bottom) as a function of collision centrality for $pp$ and Pb+Pb collisions.}
\label{fig:dijet_imbalance}
\end{center}
\end{figure}

The most striking observation by both experiments is the large centrality-dependent increase of the imbalance in the energy of the two jets, as measured in the calorimeters (See Fig.~\ref{fig:dijet_imbalance}). While their energies were very different, the two jets were observed to be very close to back-to-back in the azimuthal plane, implying little or no angular scattering of the partons during their traversal of the medium \cite{Chatrchyan:2011sx} as shown in Fig.~\ref{fig:dijet_imbalance}~\cite{Cole:2011zz}. 

\begin{figure}[ht!]
\begin{center}
   	\includegraphics[width=0.9\textwidth]{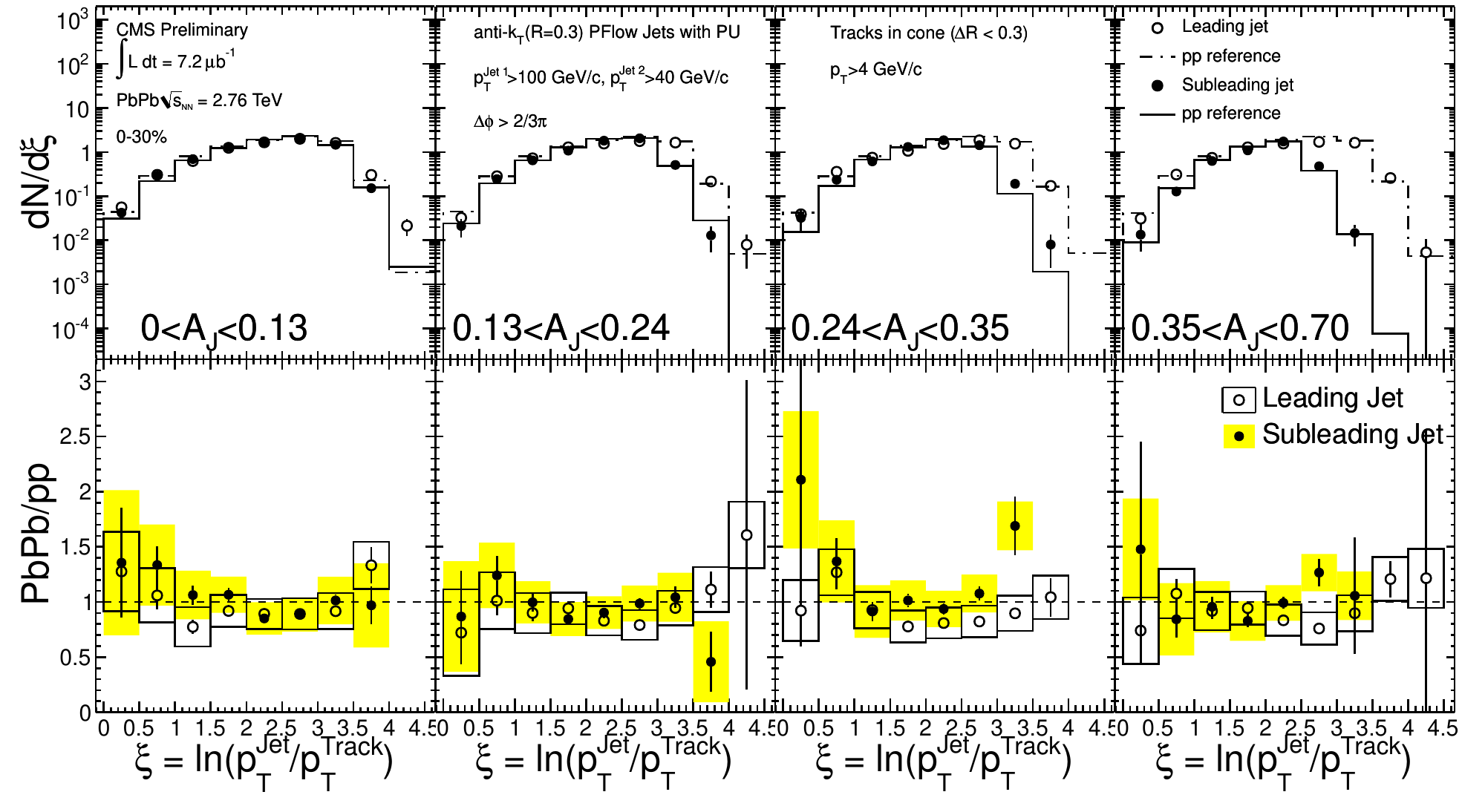}
       \caption{Ratio of Pb+Pb to $pp$ jet fragmentation function as a function of jet asymmetry $A_J$ for central Pb+Pb events.}
\label{fig:cms_fragmentation}
\end{center}
\end{figure}

The distribution of particle momenta inside jets normalized to the jet energy is the same, within experimental uncertainties, to that of jets produced in $pp$ collisions as shown in Fig.~\ref{fig:cms_fragmentation} \cite{Yilmaz:2011zz}. This suggests that most of the additional energy radiated by the leading parton inside the medium gets absorbed by the matter, and the fragments observed within the jet cone are produced outside of the medium.  Several model calculations \cite{Qin:2010mn,Lokhtin:2011qq,Young:2011qx,Vitev:2011gs}, which combine elastic and inelastic parton energy loss with deflection of radiated gluons by the medium, have been able to reproduce the increased energy asymmetry of dijets in Pb+Pb.

\begin{figure}[ht!]
\begin{center}
  	\includegraphics[width=0.75\textwidth]{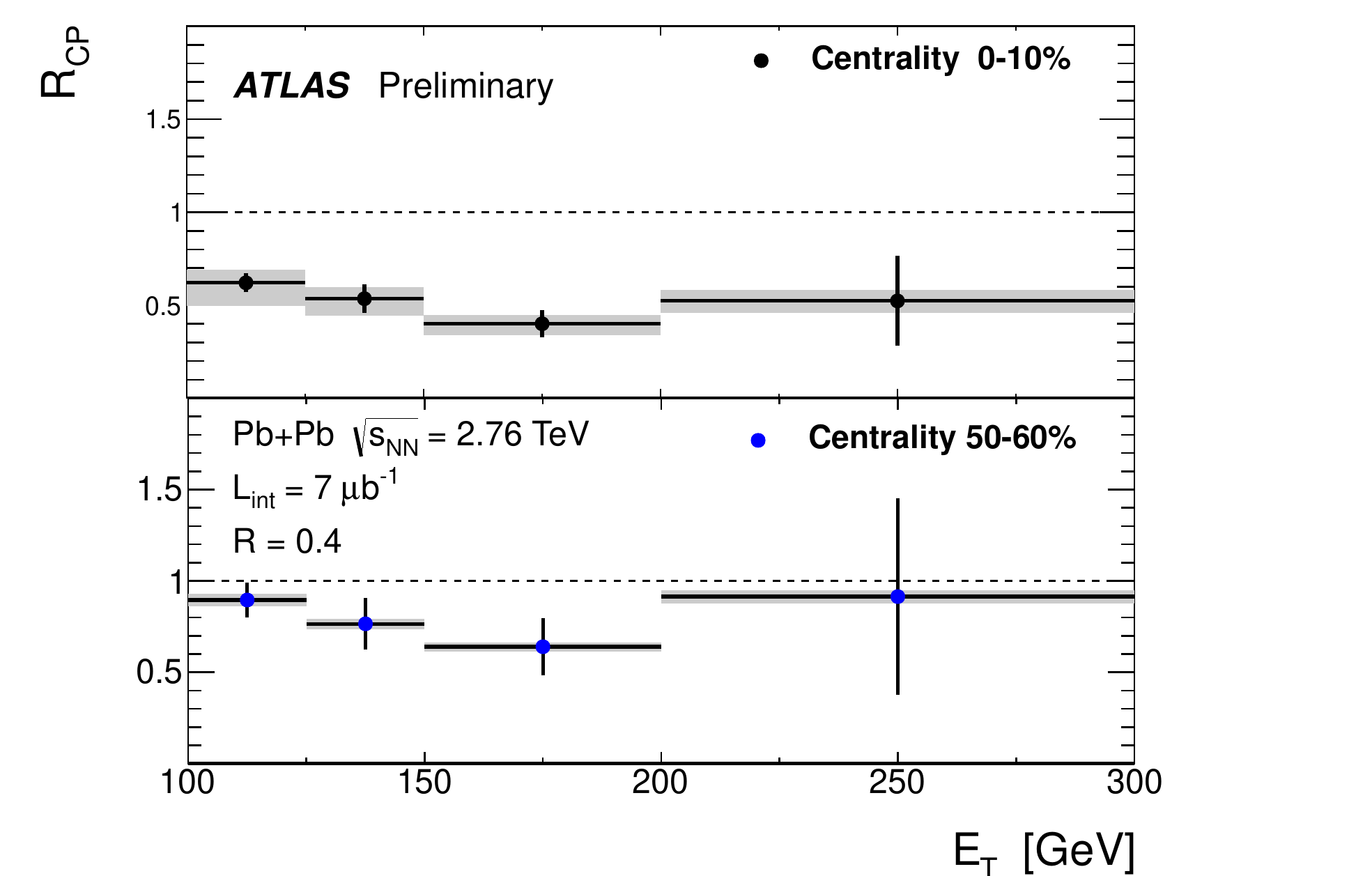}
       \caption{Jet central-to-peripheral event suppression factor $R_{\rm CP}$ for different centralities.}
\label{fig:atlas_jets}
\end{center}
\end{figure}

The strong effect of jet quenching is confirmed by the ATLAS measurement of the jet $R_{\rm CP}$~\cite{Cole:2011zz}. The comparison of jet production cross section between the central and peripheral events indicates that for the most central events the energy loss appears to reduce the cross section by about a factor of two (see Fig.~\ref{fig:atlas_jets}). This shows that the energy lost by the leading parton is not simply redistributed within the jet cone but lost from the reconstructed jet.

To find the ``missing energy'', the calorimetric measurement in CMS was complemented by a detailed study of low $\PT$ charged particles in the tracker and by using missing $\PT$ techniques. The apparent missing energy in highly unbalanced dijet events was found among the low $\PT$ particles, predominantly with $0.5<\PT<2$ \nGc, emitted outside of the sub-leading jet cone~\cite{Chatrchyan:2011sx}.

\subsection{Quarkonium suppression}

Heavy quarkonia are important probes of the quark gluon plasma since they are produced early in the collision and their survival is affected by the surrounding medium~\cite{Rapp:2009my}. The bound states of charm and bottom quarks are predicted to be suppressed in heavy ion collisions in comparison with $pp$, primarily as a consequence of deconfinment (``melting'') in the QGP~\cite{Matsui:1986dk}. The magnitude of the suppression for different quarkonium states should depend on their binding energy, with strongly bound states such as the $\Upsilon$ showing less or no modification \cite{Digal:2001ue}. 

However, \Jpsi~production, the classical deconfinement signal, has confounded expectations and interpretations ever since the first nuclear suppression was measured with Oxygen beams at the SPS~\cite{Baglin:1990iv}; now attributed to cold nuclear matter effects rather than deconfinement \cite{Gerschel:1998zi}. The ''anomalous'' suppression seen later with heavier beams~\cite{Abreu:1997jh} turned out, surprisingly, to be rather similar in magnitude at SPS and RHIC. This could indicate suppression of only the high mass charmonium states $\psi$' and $\chi_c$, which populate about 40\% of the observed \Jpsi, and which should dissociate very close to or even below the critical transition temperature. Alternatively, it has been suggested that the increasing (with energy) \Jpsi~suppression is more or less balanced by enhanced production via recombination of two independently produced charm quarks~\cite{Rapp:2008tf,BraunMunzinger:2009ih}. A resolution to this puzzle may come from measuring \Jpsi~production at the LHC where coalescence effects should be stronger because of the abundant charm production, and by comparing the suppression patterns of the \Jpsi~and $\Upsilon$ families.

\begin{figure}[!t]
\begin{tabular}{cc}
\begin{minipage}{.50\textwidth}
\centerline{\includegraphics[width=1.0\textwidth]{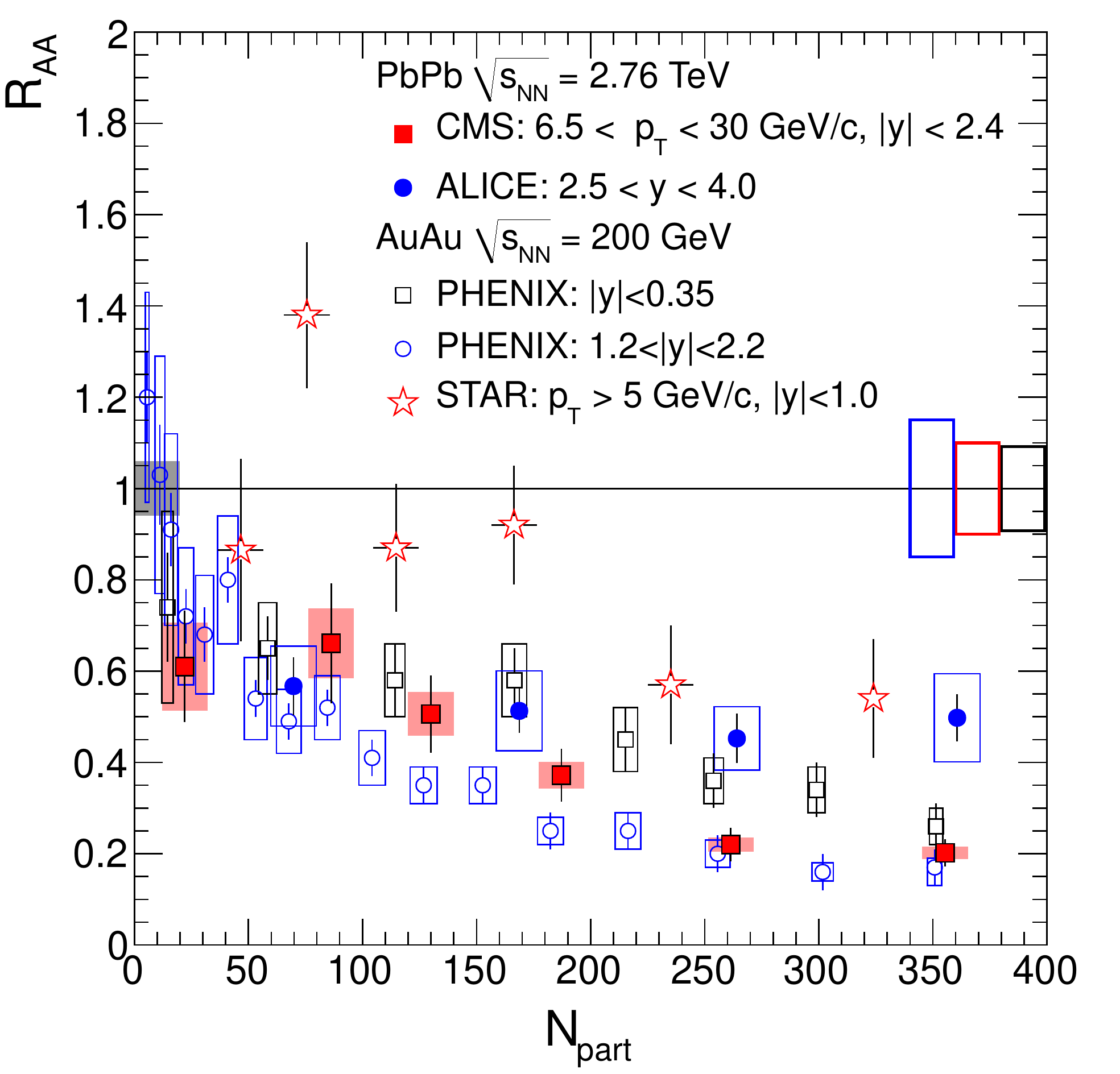}}
\end{minipage} & \begin{minipage}{.50\textwidth}
\centerline{\includegraphics[width=1.0\textwidth]{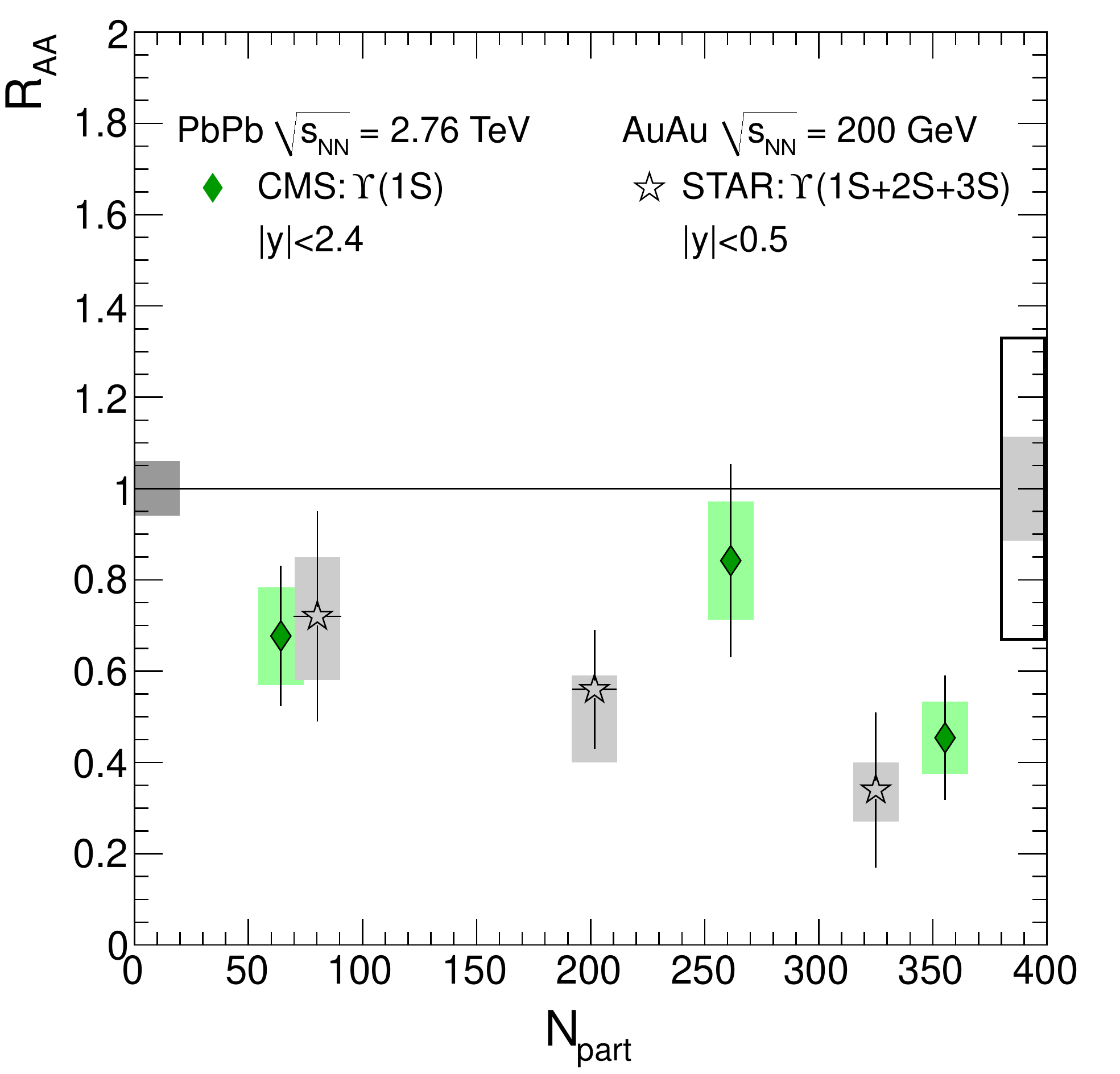}}
\end{minipage}
\end{tabular}
       \caption{Nuclear modification factor  $\RAA$ as a function of centrality for      
      $J/\psi$ (right)  and $\Upsilon$ (left).}
\label{fig:quarkonia}
\end{figure}

A compilation of first LHC results on quarkonia production~\cite{Abelev:2012rv,Chatrchyan:2012np} for both  $J/\psi$ (a) and $\Upsilon$ (b) is shown in Fig.~\ref{fig:quarkonia} as a function of centrality ($N_{part}$), together with data from RHIC~\cite{Adare:2006ns, Adare:2011yf,Reed:2011fr}.  

While errors are still large, and the overall amount of suppression at LHC remains qualitatively similar to RHIC, the detailed pattern is quite different and intriguing. The $p_T$ integrated $R_{AA}$ measured for the \Jpsi~at forward rapidity (closed circles) of about 0.5 depends very little on centrality and is almost a factor of two larger than the one measured at RHIC in central collisions, also at forward rapidity (open circles); the difference is smaller but still significant when comparing with RHIC midrapidity data (open squares). On the contrary, the high $p_T$ data at LHC (full squares), which is compatible with an independent $R_{CP}$ measurement~\cite{ATLAS:2010px}, show a stronger suppression than the high $p_T$ RHIC results (open stars). While such a pattern would be unexpected in a pure suppression scenario, it is qualitatively consistent with the recombination model, which predicts substantial regeneration effects only at low transverse momentum. The \nY~suppression (right panel) is very similar at RHIC and LHC. As only about 50\% of the observed \nY(1S) are directly produced, and the \nY(2S/3S) states seem to be more suppressed than the ground state~ \cite{Chatrchyan:2011pe}, the measured $R_{AA}$ is compatible at both RHIC and LHC with suppression of the high mass bottonium states only.

However, there are still a number of unknown aspects that must be clarified before firm conclusions can be drawn. Foremost are cold nuclear matter effects, in particular shadowing or saturation of the nuclear parton distribution functions, which for gluons are very poorly constrained at the small $x$ values relevant at LHC. Likewise, very little is known about final state absorption of quarkonia in nuclei, which may (or may not) be important at LHC. For this reason a $p$+Pb run at LHC, which will address these nuclear effects, is mandatory and anticipated for 2012.

\section{Conclusions}

The LHC has entered the field of ultra-relativistic heavy ion physics with an impressive first year performance, having benefited from many years of preparation, a strong and very complementary set of state-of-the-art detectors, and, crucially, decades of experience and progress made at previous machines. Global event properties were found to evolve smoothly from RHIC to LHC, to be qualitatively similar but quantitatively different, and to reflect the expected increases in energy density, volume, and lifetime of the hot and dense matter created in the collisions. The data have generally confirmed that also at LHC energy the bulk features imply the formation of a very strongly interacting, nearly perfect quark-gluon plasma liquid. The prominent presence of various collective flow patterns has opened a path forward to precision measurements of its material properties, including shear viscosity, equation-of-state, and sound velocity. 

Concerning bulk properties, and with the possible exception of thermal particle production, where the unexpectedly low proton yield needs confirmation and clarification, the standard reaction model established at RHIC has proven itself to be predictive and detailed enough to permit quantitative extrapolation to much higher collision energies. 

For observables where the energy reach of LHC is unique, i.~e.\ low-$x$ parton physics and hard processes, already the first low luminosity run has revealed a wealth of significant, and at times surprising results. The energy loss of high momentum partons traveling through the matter was measured with several different particle species as well as with fully reconstructed jets. While its magnitude is roughly in the expected range, the more differential analysis of parton type and momentum dependence, dijet correlations, and energy flow around the jet axis has revealed features which were not anticipated and may spur substantial refinements of our understanding and modelling of parton interactions with the hot QCD medium. 

The results on quarkonium production, where precision measurements of the $\Upsilon$ resonances are possible at LHC, in addition to those of the charmonium states, have yet to provide decisive information on the suppression mechanism.  However, intriguing hints have been found in the charmonium and bottomonium suppression pattern, suggesting significant differences to the observations at RHIC. 

The near future of heavy ion running at LHC will bring a significant increase in integrated luminosity -- already the second heavy ion run in November 2011 has increased the statistics for hard probes by more than an order of magnitude -- as well as new signals like $\gamma$--jet and $Z$--jet correlations. The $p$+Pb run planned for 2012 should not only deliver the much needed comparison data for the heavy ion program but also far extend the kinematic reach for the study of initial-state effects in the nuclear parton wave function, such as saturation and other low-$x$ physics.  And another factor of two increase in energy is in sight when LHC will reach the design value after 2014.

The combination of the on-going heavy ion program at RHIC, which includes high-luminosity measurements with upgraded detectors, the new higher energy data from LHC, and the demonstrated solid foundation in theory promises rapid progress in a quantitative understanding of the properties of the quark-gluon plasma.

\section*{Acknowledgments}

The authors thank the ALICE, ATLAS and CMS collaborations for providing the preliminary data used in this review; U.~Heinz, C.~Shen and R.~J.~Fries for  the model calculations used in Figs.~2a, 3a 6a (UH, CS) and Fig.~3a (RJF); M.~Floris, M.~Poghosyan and R.~Shahoyan for help in preparing data and figures in chapters 3 and 4;  T.~Dahms, Y.~Kim, P.~Steinberg and A.~Yoon  for preparing data and figures in chapter 5; A.~Majumder, T.~Renk, I.~Vitev and X.-N.~Wang for providing data files with the results of their calculations for Fig.~7; and M.~Floris, P.~Steinberg and S.~Voloshin for careful reading of the manuscript and helpful discussions. This work was supported in part by grants from the US DOE Office of Science (DE-FG02-05ER41367 and DE-FG02-94ER40818).


\end{document}